\documentclass[twocolumn]{autart}

\usepackage[utf8]{inputenc}
\usepackage[T1]{fontenc}
\usepackage{textcomp}

\usepackage{amsfonts}
\usepackage[tbtags]{amsmath}
\usepackage{amssymb}
\usepackage{bm}
\usepackage{mathrsfs}
\usepackage{cases}

\usepackage[colon,round,sort&compress]{natbib}

\usepackage{epsfig}
\usepackage{epsf}
\usepackage{epstopdf}
\usepackage{graphicx}
\usepackage{float}
\usepackage{subfigure}
\usepackage{cuted,lipsum,mwe}
\usepackage{multirow}
\usepackage[table]{xcolor}
\usepackage{diagbox}
\usepackage{booktabs}
\usepackage{array}
\usepackage{makecell}

\usepackage{algorithm}
\usepackage{algorithmic}
\usepackage{enumitem}

\usepackage{xparse}
\usepackage{etoolbox}

\newtheorem{definition}{Definition}
\newtheorem{lemma}{Lemma}

\newtheorem{remark}{Remark}

\newtheorem{theorem}{Theorem}
\newtheorem{corollary}{Corollary}

\newtheorem{proposition}{Proposition}

\renewcommand\arraystretch{1.2}
\newcolumntype{C}{>{\centering\arraybackslash}m{1.2cm}}

\NewDocumentCommand{\bluecitep}{o m}{%
  \begingroup
  \color{blue}%
  \IfNoValueTF{#1}
    {\citep{#2}}
    {\citep[#1]{#2}}%
  \endgroup
}

\NewDocumentCommand{\bluecitet}{o m}{%
  \begingroup
  \color{blue}%
  \IfNoValueTF{#1}
    {\citet{#2}}
    {\citet[#1]{#2}}%
  \endgroup
}

\newcommand{\newreferencekeys}{}

\let\originalbibitem\bibitem

\RenewDocumentCommand{\bibitem}{o m}{%
  \IfNoValueTF{#1}
    {\originalbibitem{#2}}
    {\originalbibitem[#1]{#2}}%
  \ifinlist{#2}{\newreferencekeys}
    {\color{blue}}
    {\color{black}}%
}

\begin{document}

\begin{frontmatter}

\title{Stable and fair benefit allocation in mixed-energy truck platooning: A coalitional game approach} 
\vspace{-5pt}

\thanks[footnoteinfo]
{This research was supported in part by NSF under Grants CNS-2401007, CMMI-2348381, IIS-2415478, in part by MathWorks, in part by the National Natural Science Foundation of China under Grant 62603450, and in part by the Start-Up Fund of Shanghai Jiao Tong University under Grant WL220203203. Corresponding author: Xinfeng Ru.}

\author[SJTU]{Ting Bai}\ead{tingbai@sjtu.edu.cn}, \author[ECUST]{Xinfeng Ru}\ead{xfru@ecust.edu.cn}, \author[KTH]{Karl H. Johansson}\ead{kallej@kth.se}, \author[KTH]{{Jonas M{\aa}rtensson}}\ead{jonas1@kth.se}, and \author[Cornell]{Andreas A. Malikopoulos}\ead{amaliko@cornell.edu}

\address[SJTU]{School of Automation and Intelligent Sensing, Shanghai Jiao Tong University, China}
\address[ECUST]{School of Information Science and Engineering, East China University of Science and Technology, China}
\address[KTH]{Division of Decision and Control Systems, KTH Royal Institute of Technology, Sweden}
\address[Cornell]{Information and Decision Science Lab, School of Civil \& Environmental Engineering, Cornell University, USA}  
\vspace{-20pt}


\begin{keyword}                           
Mixed-energy truck platoon; coalitional game; core stability; Shapley value.               
\end{keyword}                             

\begin{abstract}                           
This paper addresses the benefit allocation in a mixed-energy truck platoon composed of fuel-powered and electric trucks. The interactions among trucks during platoon formation are modeled as a coalitional game with transferable utility. We first design a stable payoff allocation scheme that accounts for truck heterogeneity in energy savings and platoon roles (leader or follower), establishing core-stability conditions to ensure that no subset of trucks has an incentive to deviate for greater benefit. To enhance payoff fairness, we then propose a closed-form, Shapley value-based allocation approach that is computationally efficient and independent of the platoon size. Sufficient conditions under which the allocation is both fair and core-stable are provided. In scenarios where the Shapley value falls outside the core, we develop an alternative allocation based on the stable payoff that minimizes the mean relative deviation from the Shapley value while preserving core stability. This deviation is further proved to be upper-bounded by $1$, showing a favorable trade-off between stability and fairness. Finally, extensive numerical studies validate the theoretical results and demonstrate the effectiveness of the proposed framework in facilitating stable, equitable, and sustainable cooperation in mixed-energy truck platooning. 
\vspace{-5pt}
\end{abstract}
\end{frontmatter}

\section{Introduction}\label{Section 1}
Truck platooning has emerged as a promising technique in road transportation~\citep{mahbub2023_automatica,Kumaravel:2021uk,11423204}. Advanced from vehicle-to-vehicle communication and automated driving systems, platooning enables trucks to form coordinated convoys with small inter-vehicle distances, known as \emph{platoons}. Due to the reduced aerodynamic drag experienced by trailing vehicles, platooning leads to energy savings of up to $20\%$ and reduces associated greenhouse gas emissions significantly~\citep{liang2015heavy}. Compared to driving independently, joining platoons yields notable benefits, including improving road capacity, mitigating traffic congestion, reducing drivers' workload, and enhancing driving safety, to name only a few; see~\cite{jo2019benefits,Beaver2021Constraint-DrivenStudy,10209062,Kumaravel:2021wi}. These advantages make platooning an attractive strategy for freight carriers seeking to improve transportation efficiency, save operational costs, and minimize their environmental footprint~\citep{mahbub2021_platoonMixed}. 
\vspace{-4pt}

Extensive experimental and field tests have shown that trailing trucks in a platoon achieve energy savings ranging from $10\%$ to $18\%$, depending on the inter-vehicle gap~\citep{tsugawa2016review}. The lead truck, however, often benefits far less, with fuel consumption reductions below $2\%$ when the gap exceeds $15$ meters~\citep{bishop2017evaluation}. This inherent imbalance in platooning benefits between the lead and following trucks creates asymmetric incentives for participation, especially for those operated by different carriers, prioritizing their own gains over collective efficiency. To enable effective cooperation in platoon formation and maximize both individual and system-wide benefits, equitable allocation schemes that redistribute platooning savings among all participating trucks are required. 
\vspace{-4pt}

Conventional platoons typically consist of uniformly fuel-powered trucks (FPTs) with similar sizes and loads, where each following vehicle achieves comparable fuel savings. In such platoons, a simple yet effective benefit allocation approach is to evenly distribute the total fuel savings of the following trucks among all participants, as widely adopted in practice; see~\cite{9435129,9993403,ZENG2025106452}. This method compensates the lead truck, whose direct fuel savings from platooning are negligible~\citep{van2017fuel}, with an equal share as that allocated to each follower, thus promoting fairness among all platoon members. While the even-split mechanism proves effective in \emph{homogeneous} platoons, it falls short in mixed traffic scenarios involving diverse truck types. In such cases, imbalances arise not only between the lead and following trucks but also among the followers themselves.

The freight transportation sector today is undergoing a fundamental transition toward electrification to overcome escalating energy shortages and mitigate the growing impacts of climate change~\citep{bakker2025strategic}. As part of this shift, an increasing number of logistics providers are integrating electric trucks (ETs) into their fleets to comply with tightening emissions regulations and achieve sustainability goals. With the rising adoption of ETs, both FPTs and ETs appear in platoons, leading to \emph{heterogeneous} energy profiles. While platooning remains advantageous in mixed-energy truck compositions, the magnitude and nature of the benefits, such as energy consumption patterns and cost savings, vary significantly across vehicle types. This heterogeneity poses new challenges in fairly and stably allocating the platooning benefits among platoon participants. 

This paper investigates the benefit allocation problem in mixed-energy truck platooning, where fuel-powered and electric trucks coexist within a platoon. Each truck is assumed to be owned by a different carrier, and the objective is to achieve a \emph{stable} and \emph{fair} platooning benefit distribution among all participants. In practice, a well-designed stable allocation scheme prevents the formation of short-lived or fragmented platoons (i.e., coalitions). By ensuring that no subset of trucks has a profitable incentive to deviate, stable benefit allocation eliminates trucks' concerns about coalition collapse, thereby fostering reliable and sustained cooperation among trucks. This is particularly important in repeated or multi-stage interactions. At the same time, fairness in benefit allocation ensures that the value created by a cooperative platoon is shared equitably, so that every truck, despite its type (FPT or ET) and role (leader or follower), perceives the outcome as just and satisfactory. A fair allocation reduces the risk of internal conflict, which in turn enhances trucks' willingness to cooperate and prompts enduring stability. Without a transparent, stable, and fair allocation mechanism, trucks that consistently bear higher costs or receive disproportionately lower benefits may opt out of cooperation, leading to decreasing participation rates, waste of platooning opportunities, and inefficiencies in system performance; see~\cite{BOUCHERY202215}. Addressing these challenges is vital to enabling cohesive platoon formations and fully unlocking the potential of truck platooning.  
\vspace{-4pt}

Considerable studies have proposed cost-sharing or incentive schemes for homogeneous FPT platoons. For instance, \citet{farokhi2013game} and \citet{johansson2019game} presented benefit distribution models to address the platoon matching problem, where vehicles are modeled as strategic agents that independently optimize their departure times to maximize their platooning gains individually. The interactions among trucks are formulated as a non-cooperative game, with the Nash equilibrium adopted as the solution concept. Accordingly, the research effort in \cite{sun2019behaviorally} addressed the same issue by first maximizing the profit of all vehicles from platooning and then allocating the benefits in a way that no vehicles have incentives to deviate from their assigned platoon. However, most existing approaches do not incorporate heterogeneous vehicle settings, where fairness and stability are more difficult to guarantee due to the interplay between energy heterogeneity and vehicle roles in the platoon. More recently, \cite{chen2023cost} studied the cost allocation among multiple carriers, where an alliance operator collects requests and coordinates autonomous trucks to form platoons while minimizing the total cost. While the fuel consumption cost is heterogeneous depending on the truck loading and engine attributes, the proposed allocation models can only achieve a trade-off between stability and efficiency with \emph{approximate} core-stability guarantees. 
\vspace{-4pt}

In this paper, we address the challenges of benefit allocation in mixed-energy truck platooning, where trucks are either fuel-powered or electric. We study the problem in a classical hub-based platoon formation framework, where trucks form platoons at logistics hubs and travel together along shared routes. Each truck, representing a self-interested carrier, aims to maximize its own platooning benefits by coordinating with others during platoon formation. We model the interactions among trucks as a coalitional game with transferable utility and propose benefit allocation schemes that ensure both core stability and fairness in the payoff distribution. Our approach accounts not only for the profit imbalance between the lead and following trucks but also for disparities between different truck types. The developed benefit-sharing mechanism is designed to serve as a binding agreement among participating trucks, facilitating stable and sustainable platoon formations across carriers. The main contributions of this paper are as follows:
\vspace{-5pt}
\begin{itemize}
\item We model the strategic interactions among trucks as a platoon coalitional game and demonstrate that the grand coalition constitutes the optimal coalition structure, resulting in the highest total profit and benefiting every platoon participant (Lemma~\ref{Lemma1}).
\vspace{2.5pt}

\item We propose a payoff allocation scheme to distribute the total platooning benefit among all participants in the optimal coalition structure, accounting for heterogeneity in both role (leader or follower) and truck type (FPT or ET), while ensuring core stability of the allocation (Theorem~\ref{Theorem1}).
\vspace{2.5pt}

\item We present a closed-form Shapley value-based payoff allocation (Proposition~\ref{Proposition1}), providing an efficient and fair approach to allocate the total platooning benefits among all participants. Sufficient conditions under which the Shapley value lies within the core of the game are identified (Theorem~\ref{Theorem2}). 
\vspace{2.5pt}

\item We develop a payoff allocation scheme that balances fairness and stability in scenarios where the Shapley value falls outside the core. The allocation minimizes the mean relative deviation of individual payoffs from the Shapley values, with the deviation upper-bounded by $1$, achieving minimal unfairness while preserving core stability (Theorem~\ref{Theorem3}).
\end{itemize}
\vspace{-10pt}

The proposed approach provides a novel framework for fair and stable distribution of platooning benefits among mixed-energy truck participants, thus facilitating efficient platoon formation and promoting sustained cooperation. Notably, the widely applied even-split allocation mechanism in homogeneous truck platooning emerges as a special case of our method, satisfying the conditions for both core stability and fairness (Corollary~\ref{Corollary1} and Proposition~\ref{Proposition2}). Finally, extensive numerical studies are conducted to validate the theoretical results. 
\vspace{-10pt}

The rest of the paper is organized as follows. Section~\ref{Section 2} introduces the mixed-energy truck platooning problem and the modeling of the coalitional game. Section~\ref{Section 3} presents a payoff allocation scheme ensuring core stability. To enhance fairness, Section~\ref{Section 4} proposes a closed-form Shapley value-based allocation and identifies conditions under which the Shapley value lies within the core. Section~\ref{Section 5} deals with cases where the Shapley value is not core-stable, and Section~\ref{Section 6} validates the key theoretical findings with numerical examples. Eventually, concluding remarks and potential directions for future research are discussed in Section~\ref{Section 7}.
\vspace{-10pt}

\textit{Notation:} Let $\mathbb{R}$ and $\mathbb{R}_{+}$ be the sets of real and positive real numbers, respectively, and $\mathbb{R}^n$ the $n$-dimensional Euclidean space. Let $\mathbb{N}$ denote the set of non-negative integers. The transpose of a vector is denoted by $(\cdot)^{\top}$, and the cardinality of a set $\mathcal{S}$, i.e., the number of elements the set contains, is denoted as $|\mathcal{S}|$. The binomial coefficient $\binom{n}{m}$, where $n,m\!\in\!{\mathbb{N}}$ and $m\!\leq\!{n}$, represents the number of distinct combinations of $m$ elements selected from a set of $n$ elements. The factorial of $n$, denoted as $n!$, is the product of all positive integers less than or equal to $n$, with the convention $0!\!=\!1$.
\vspace{-3pt}

\section{Problem formulation}\label{Section 2}
\subsection{Mixed-energy truck platooning}\label{Section2.1}
\begin{figure}[t]
\centering
\includegraphics[scale=0.48]{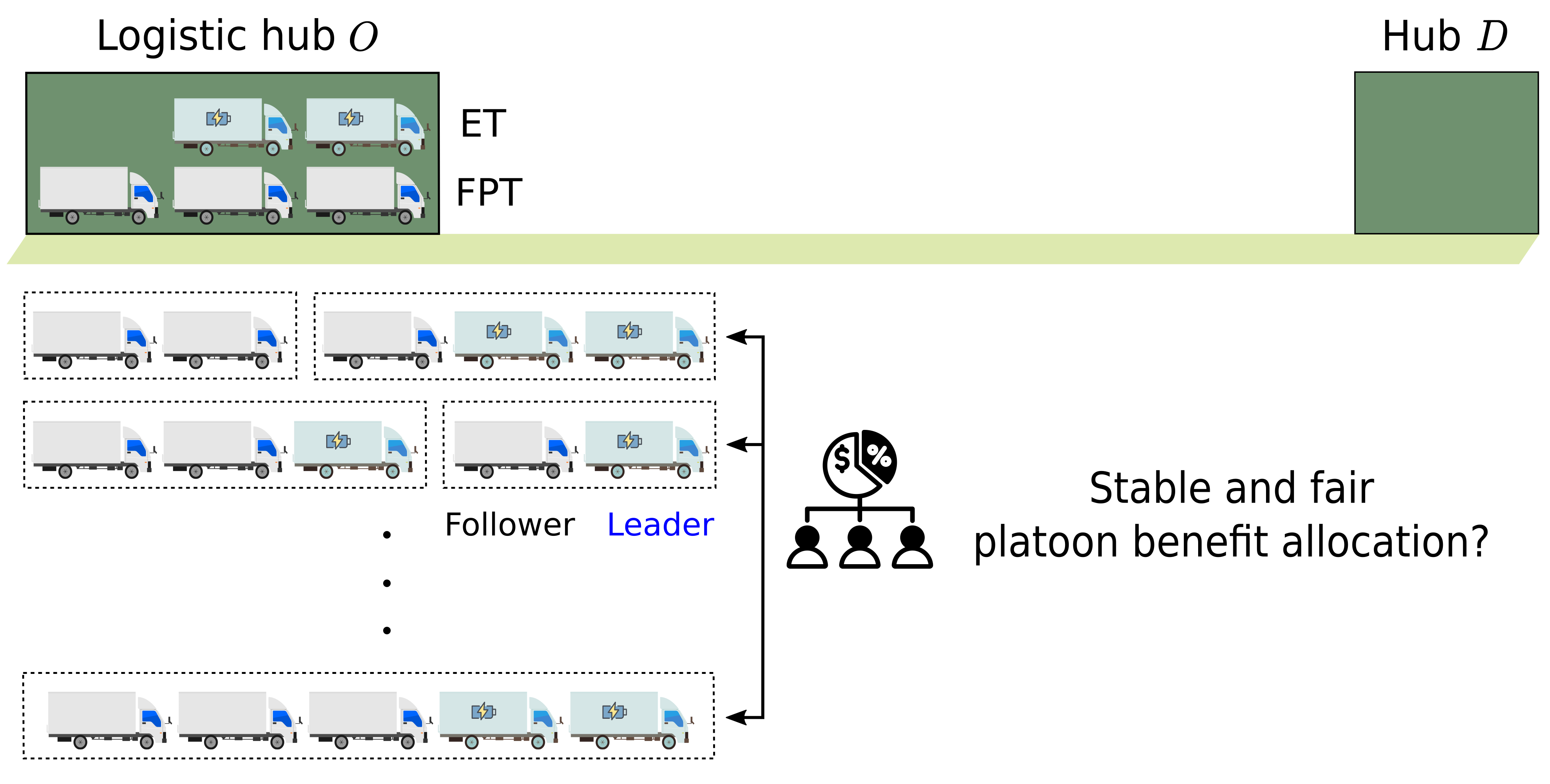}
\vspace{-5pt}
\caption{Illustration of the benefit allocation in mixed-energy truck platooning. The potential platoon formation configurations are shown by blocks with dotted lines.}
\label{Fig.1}
\end{figure}

We consider a set of $N$ trucks at an origin logistic hub $O$, each planning to travel to the destination hub $D$, as illustrated in Fig.~\ref{Fig.1}. The set of trucks is denoted by $\mathcal{N}\!:=\!\{1,2,\dots,N\}$, consisting of $N_f\!=\!|\mathcal{N}_f|$ FPTs and $N_e\!=\!|\mathcal{N}_e|$ ETs, with $N\!\in\!\mathbb{N}$, $N\!\geq\!{2}$, and $\mathcal{N}\!\!=\!\mathcal{N}_f\!\cup\!{\mathcal{N}_e}$. For technical and safety reasons~\citep{boysen2018identical}, the platoon size, i.e., the number of trucks allowed to form a single platoon, is limited to $M\!\in\!\mathbb{N}$. As a result, the number of trucks at the origin hub satisfies $2\!\leq\!{N}\!\leq\!{M}$. 
\vspace{-3pt}

Each truck is assumed to be operated by a distinct carrier focused on maximizing its own platooning benefit. Suppose that every truck is equipped with sufficient fuel or battery capacity to independently complete the journey from hub $O$ to hub $D$. Nevertheless, to improve energy efficiency, save operational costs, and reduce carbon emissions, trucks seek to form platoons with others at the origin hub. In each platoon, the lead truck is assumed to receive zero direct benefit from platooning~\citep{liang2015heavy,larsen2019hub}, while each following truck obtains a benefit that depends on its type. The heterogeneity in the follower truck benefits is characterized by $\epsilon_f,\epsilon_e\!\in\!{\mathbb{R}_{+}}$, where $\epsilon_f$ and $\epsilon_e$ denote the monetary savings per unit travel distance for a fuel-powered and an electric follower, respectively, over a hub-to-hub segment under certain operating conditions. In the model considered, $\epsilon_f$ and $\epsilon_e$ are exogenous parameters attainable prior to coalition formation and can be estimated from experimental or field data. Typically, considering different energy prices and consumption, there is $\epsilon_e\!<\!\epsilon_f$. For presentation simplicity, the distance between the two hubs is normalized to~$1$. In this paper, we aim to address the following questions:
\vspace{-8pt}
\begin{itemize}
\item [(i)] What is the optimal platoon formation structure that maximizes the collective platooning benefits of all participating trucks?
\vspace{2pt}
\item [(ii)] How to allocate the maximized platooning benefits among the participating trucks in a way that is both fair and stable? 
\end{itemize}
\vspace{-3pt}

Forming platoons across different carriers requires cooperation among trucks, building upon a \emph{binding agreement} that specifies how the resulting benefits will be distributed among participants in an acceptable and, ideally, fair manner. Addressing question (i) determines the optimal platoon configuration, which identifies the most effective scope of cooperation that maximizes the collective platooning benefit, while addressing question (ii) ensures stability and fairness of the agreement by fully taking into account each truck's contribution to the successful formation of the platoon. 
\vspace{-5pt}

\begin{remark}\label{Remark0}
The leader’s direct platooning benefit is assumed to be zero in this paper. This is a reasonable assumption for truck platoons operating with relatively large inter-vehicle gaps, where the leader’s benefit is typically much smaller than that of the followers and is thus often neglected in platoon coordination models~\citep{9435129,van2017fuel}. The proposed framework can be extended to settings with a nonzero leader benefit by redefining the coalition worth function and re-deriving the allocation and core-stability conditions. A small leader benefit would mainly alter the analytical thresholds and closed-form expressions, without changing the essential problem structure and the underlying stability-fairness trade-off. To retain a transparent and tractable closed-form analysis, we maintain the zero leader-benefit assumption in this work.
\end{remark}

\begin{remark}\label{Remark1}
    To ensure clarity and focus on the benefit allocation schemes, this paper assumes that all trucks at the origin hub $O$ are ready to depart without incurring additional waiting times. The decision thus involves only forming optimal platoons under a transparent binding agreement that is both fair and stable, guaranteed by the proposed benefit allocation approach. It is worth noting that our framework can be readily extended to multi-hub coordination under other platooning constraints, such as departure time scheduling subject to trucks' charging or waiting requirements; see~\cite{johansson2019game,sun2019behaviorally,10932686n}.   
\end{remark}

\subsection{Coalitional game formulation}\label{Section2.2}
In this subsection, we model the strategic interactions among trucks as a coalitional game. To this end, we first introduce relevant concepts from cooperative game theory. For further background on game theory, we refer readers to~\cite{fudenberg2026game,peleg2007introduction,peters2008game,gibbons1992game}.

\begin{definition}[Coalitional game] \label{Def1}
A coalitional game with transferable utility is a pair $(\mathcal{N},v)$, where
\vspace{-8pt}

\noindent $(1)$ $\mathcal{N}\!=\!\{1,2,\dots,N\}$, $N\!\in\!\mathbb{N}$, is a finite set of players;\\
\noindent $(2)$ $v\!:\!2^{\mathcal{N}}\!\! \to \!\mathbb{R}$ is a characteristic function, which assigns a real value $v(\mathcal{S})$ to each coalition $\mathcal{S}\!\subseteq\!{\mathcal{N}}$ with $v(\emptyset)\!=\!0$. The game has transferable utility if the value $v(\mathcal{S})$ can be freely distributed among members of the coalition $\mathcal{S}$.
\end{definition}

Notice that the characteristic function $v$ maps each subset of $\mathcal{N}$ into a real number. The set of all such subsets (i.e., the powerset of $\mathcal{N}$) is denoted by $2^{\mathcal{N}}$, including the empty set $\emptyset$. 

\begin{definition}[Outcome]\label{Def2}
An outcome of a coalitional game $G\!=\!(\mathcal{N},v)$ is a pair $(\mathcal{P},x)$, where 
\vspace{-8pt}

\noindent $(1)$ $\mathcal{P}\!=\!\{\mathcal{S}_1,\dots,\mathcal{S}_P\}$ with $\mathcal{S}_m\!\neq\!{\emptyset}$, $m\!=\!1,\dots,P$, is a coalition structure, i.e., a partition of $\mathcal{N}$, such that\\
\noindent $(1a)$ $\mathcal{S}_m\!\cap\!{\mathcal{S}_\ell}\!=\!\emptyset$, $\forall{\mathcal{S}_m},\mathcal{S}_\ell\!\in\!\mathcal{P}$, $m\!\neq\!{\ell}$ (disjoint coalitions); \\
\noindent $(1b)$ $\bigcup_{m=1}^{P}\!\mathcal{S}_m\!=\!\mathcal{N}$ (complete coverage).\\
\noindent $(2)$ $x\!=\![x_1,\dots,x_N]^{\top}\!\!\in\!{\mathbb{R}^N}$ is a payoff vector satisfying\\
\noindent $(2a)$ $x_i\!\geq\!{v(\{i\})}$, $\forall{i}\!\in\!{\mathcal{N}}$ (individual rationality);\\
\noindent $(2b)$ $\sum_{i\in{\mathcal{S}_m}}\!\!\!x_i\!=\!v(\mathcal{S}_m)$, $\forall{\mathcal{S}_m}\!\in\!\mathcal{P}$ (coalitional efficiency). 
\end{definition}

The subsets $\mathcal{S}_1,\dots,\mathcal{S}_P\!\subseteq\!{\mathcal{N}}$ are called coalitions, and the set $\mathcal{N}$ is known as \emph{grand coalition}. For any coalition $\mathcal{S}\!\subseteq\!{\mathcal{N}}$, including the empty set $\emptyset$, $v(\mathcal{S})$ denotes the total value (or payoff) that the players of $\mathcal{S}$ can achieve through cooperation in the game $G\!=\!(\mathcal{N},v)$.

\begin{definition}
\cite[p.~124]{peters2008game}
\label{Def3}
The core of a coalitional game $G\!=\!(\mathcal{N},v)$, denoted by $\mathrm{Core}(G)$, is the set of payoff vectors $x\!=\![x_1,\dots,x_N]^{\top}\!\!\in\!\mathbb{R}^N$ satisfying 
\vspace{-6pt}
\begin{eqnarray}
&&\!\mathrm{Core}(G)\!\nonumber\\
&&=\!\Big\{\!x\!\in\!\mathbb{R}^N \big |\!\sum_{i\in\mathcal{N}}\!x_i\!=\! v(\mathcal{N}) \text{and}\sum_{i\in\mathcal{S}}\!x_i\!\geq\! v(\mathcal{S}),\forall\mathcal{S}\!\subseteq\!\mathcal{N}\Big\}.\!\label{Equ.1}
\end{eqnarray}
\end{definition}
\vspace{-10pt}
The core of a game is the set of payoff allocations in which no coalition has an incentive to deviate from the grand coalition. The \emph{efficiency} condition, i.e., $\sum_{i\in\mathcal{N}}\!x_i\!=\!v(\mathcal{N})$, ensures that the total value created by the grand coalition $\mathcal{N}$ is fully allocated among all players. The \emph{coalitional rationality} condition, i.e., $\sum_{i\in{\mathcal{S}}}\!x_i\!\geq\!{v(\mathcal{S})},\forall{\mathcal{S}}\!\subseteq\!{\mathcal{N}}$, guarantees that no subset of players can achieve a higher collective payoff by breaking away to form new coalitions. A payoff vector $x$ lies in the core if and only if it is both efficient and coalitionally rational.
\vspace{-5pt}

We now proceed to present the coalitional game formulation of the mixed-energy truck platooning problem. Let $G\!=\!(\mathcal{N},v)$ denote the \emph{platoon coalitional game} with transferable utility at the origin hub $O$, where $\mathcal{N}\!=\!\{1,2,\dots,N\}$, with $2\!\leq\!{N}\!\leq\!{M}$, represents the set of trucks (i.e., players). Recall that $\mathcal{N}\!=\!\mathcal{N}_f\!\cup\!{\mathcal{N}_e}$, where $\mathcal{N}_f$ and $\mathcal{N}_e$ denote the sets of FPTs and ETs, respectively. The platooning benefit is monetized and thus can be freely transferred among participating trucks. For any coalition $\mathcal{S}\!\subseteq\!{\mathcal{N}}$, the characteristic function $v(\mathcal{S})$ represents the maximum total platooning benefit achievable by trucks in $\mathcal{S}$ through optimal leader selection. In homogeneous platoons comprising identical trucks, any leader selection yields the same total benefit. However, in mixed-energy truck platoons, where follower gains vary by truck type, strategic leader selection becomes essential to maximize the total platooning benefit.     

Specifically, for any coalition $\mathcal{S}\!\subseteq\!{\mathcal{N}}$, let $N^S\!\!=\!|\mathcal{S}|$ be the total number of trucks in the coalition. The numbers of FPTs and ETs in $\mathcal{S}$ are denoted by $N_f^S\!=\!|\mathcal{N}_f^{\mathcal{S}}|\!=\!|\mathcal{S}\!\cap\!{\mathcal{N}_f}|$ and $N_e^S\!=\!|\mathcal{N}_e^{\mathcal{S}}|\!=\!|\mathcal{S}\!\cap\!{\mathcal{N}_e}|$, respectively. By definition, it satisfies $N^S\!=\!N_f^S+N_e^S$. The characteristic function $v(\mathcal{S})$, with $\mathcal{S}\!\subseteq\!{\mathcal{N}}$, is defined as
\vspace{-8pt}
\begin{eqnarray}
v(\mathcal{S})&&\!\!:=\!\max_{L\in\mathcal{S}}v(\mathcal{S},L)\!\nonumber\\
&&\!=\!\!\begin{cases}
	\!0,& \!\!\!\text{if~$N^S\!\!=\!0$,} \\
	\!\epsilon_e(N_e^S\!\!-\!1)\!+\!\epsilon_fN_f^S,& \!\!\text{if~$N^S\!\!\geq\!1$,$N_e^S\!\!\geq\!1$,$L\!\in\!\mathcal{N}_e^{\mathcal{S}}$\!,}\\
    \!\epsilon_f(N_f^S\!\!-\!1),& \!\!\text{if~$N^S\!\!\geq\!1$,$N_e^S\!\!=\!0$,$L\!\in\!\mathcal{N}_f^{\mathcal{S}}$.}
	\end{cases}\label{Equ.2}
\end{eqnarray}
Here, $L$ denotes the leader of the platoon formed by $\mathcal{S}$, and $v(\mathcal{S})$ represents the total platooning benefit achieved by all following trucks in $\mathcal{S}$, under the assumption that the leader contributes no savings. The first case in \eqref{Equ.2} corresponds to $\mathcal{S}\!=\!\emptyset$. As $\epsilon_e\!<\!\epsilon_f$, assigning an ET as the leader maximizes the total platooning benefit if $N_e^S\!\!\geq\!{1}$. This is captured by the second case of \eqref{Equ.2}, which also includes the homogeneous platooning case that consists of only ETs (i.e., when $N^S\!=\!N_e^S$). Note that $\epsilon_e\!<\!\epsilon_f$ may not hold under different operating conditions. Nevertheless, the proposed framework remains applicable under the reversed ordering by selecting the leader from the truck type with the lower follower-saving coefficient. The third case characterizes scenarios where all trucks in $\mathcal{S}$ are fuel-powered, namely, when $N_e^S\!\!=\!0$. The second and third cases of \eqref{Equ.2} also encompass the situation where $N^S\!=\!1$, corresponding to a truck traveling alone without forming platoons, resulting in zero platooning benefits with $v(\mathcal{S})\!=\!0$.
\vspace{-5pt}

Based on the platoon coalitional game formulated above, the questions raised earlier can be reframed as follows:
\vspace{-8pt}
\begin{itemize}
\item [(i)] What is the optimal coalition structure $\mathcal{P}$ in the platoon coalitional game $G\!=\!(\mathcal{N},v)$ that maximizes the collective payoff?
\vspace{2pt}
\item [(ii)] Under the optimal coalition structure, how can one determine a payoff vector $x$ that is both core-stable and fair, accounting for each truck’s distinct role and heterogeneous contribution?
\end{itemize} 
\vspace{-5pt}

\section{Stable payoff allocation}\label{Section 3}
In this section, we investigate a fundamental property of the platoon coalitional game $G\!=\!(\mathcal{N},v)$. Taking this as a basis, we propose a payoff allocation scheme that ensures core stability. 

\subsection{Superadditive property of game $G$}\label{Section3.1}
As presented in Definition~\ref{Def2}, an outcome of a coalitional game is comprised of a coalition structure $\mathcal{P}$ and a payoff vector $x$ that satisfy certain conditions. To find the optimal coalition structure that maximizes the overall payoff, we first explore key properties of the game $G$. Towards this goal, the following concept is introduced. 

\begin{definition} \cite[p.~10]{peleg2007introduction}
\label{Def4}
A coalitional game $G\!=\!(\mathcal{N},v)$ is superadditive if $v(\mathcal{S}\!\cup\!{\mathcal{T}})\!\geq\!{v(\mathcal{S})\!+\!v(\mathcal{T})}$ for all coalitions $\mathcal{S},\mathcal{T}\!\subseteq\!{\mathcal{N}}$ with $\mathcal{S}\cap{\mathcal{T}}\!=\emptyset$.
\end{definition}

Superadditivity in a coalitional game implies that cooperation between disjoint coalitions leads to greater value. That is, when two disjoint coalitions merge, the value of their union is at least as large as the sum of their individual values. This property incentivizes players to form larger coalitions, as collaboration never decreases the total value. For the mixed-energy truck platooning problem modeled by $G$, we have the following result.

\begin{lemma}\label{Lemma1}
The platoon coalitional game $G\!=\!(\mathcal{N},v)$ with $v$ defined in \eqref{Equ.2} is superadditive.
\end{lemma}

\noindent \textbf{Proof.} Let $\mathcal{P}\!=\!\{\mathcal{S}_1,\dots,\mathcal{S}_P\}$ be a coalition structure of game $G$, satisfying $\mathcal{S}_m\!\cap\!{\mathcal{S}_{\ell}}\!=\!\emptyset$, $\forall\mathcal{S}_m,\mathcal{S}_{\ell}\!\in\!\mathcal{P}$, $m\!\neq\!{\ell}$, and $\bigcup_{m=1}^P\!\mathcal{S}_m\!=\!\mathcal{N}$. For any $\mathcal{S}_m\!\in\!{\mathcal{P}}$, let $\mathcal{N}_f^{\mathcal{S}_m}\!=\!\mathcal{S}_m\!\cap\!{\mathcal{N}_f}$ and $\mathcal{N}_e^{\mathcal{S}_m}\!=\!\mathcal{S}_m\!\cap\!{\mathcal{N}_e}$ be the sets of FPTs and ETs in $\mathcal{S}_m$, respectively. The cardinalities of the sets $\mathcal{N}_f^{\mathcal{S}_m}$, $\mathcal{N}_e^{\mathcal{S}_m}$, and $\mathcal{S}_m$ are denoted by $N_f^{S_m}$, $N_e^{S_m}$, and $N^{S_m}$, respectively. By definition, $N^{S_m}\!=\!N_f^{S_m}\!+\!N_e^{S_m}$.

For any two coalitions $\mathcal{S}_m,\mathcal{S}_{\ell}\!\in\!\mathcal{P}$, we first derive the characteristic function of their union. As $\mathcal{S}_m\!\neq\!{\emptyset}$ and $\mathcal{S}_{\ell}\!\neq\!{\emptyset}$ , by \eqref{Equ.2}, one has  
\vspace{-5pt}
\begin{eqnarray}
&&\!\!v(\mathcal{S}_m\!\cup\!{\mathcal{S}_{\ell}})=\!\max_{L\in{\mathcal{S}_m\cup{\mathcal{S}_{\ell}}}}\!v(\mathcal{S}_m\!\cup\!{\mathcal{S}_{\ell}},L)=\nonumber\\
&&\!\!\begin{cases}
\!\!\epsilon_e(N_e^{S_m}\!\!+\!N_e^{S_\ell}\!\!-\!1)\!+\!\epsilon_f(N_f^{S_m}\!\!+\!N_f^{S_\ell}),&\text{\!\!\!if~$N_e^{S_m}\!+\!N_e^{S_\ell}\!\!\geq\!{1}$,}\\ 
& \  \text{$L\!\in\!{\mathcal{N}_e^{\mathcal{S}_m}}\!\cup\!{\mathcal{N}_e^{\mathcal{S}_\ell}}$\!,}\\ 
\!\!\epsilon_f(N_f^{S_m}\!\!+\!N_f^{S_\ell}\!\!-\!1),&\text{\!\!\!if~$N_e^{S_m}\!+\!N_e^{S_\ell}\!\!=\!0$,}\\
& \  \text{$L\!\in\!{\mathcal{N}_f^{\mathcal{S}_m}}\!\cup\!{\mathcal{N}_f^{\mathcal{S}_\ell}}$\!.}\nonumber
\end{cases}
\end{eqnarray}
Given that $N_e^{S_m}\!\geq\!{0}$ and $N_e^{S_\ell}\!\geq\!{0}$, to compute the individual values of $v(\mathcal{S}_m)$ and $v(\mathcal{S}_\ell)$, we examine the following three cases:

(C1) $N_e^{S_m}\!=\!0$ and $N_e^{S_\ell}\!=\!0$:\\
In this case, $N_f^{S_m}\!\geq\!{1}$ and $N_f^{S_\ell}\!\geq\!{1}$. In line with $v(\mathcal{S})$ defined in \eqref{Equ.2}, we have 
\vspace{-5pt}
\begin{eqnarray}
v(\mathcal{S}_m)\!+\!v(\mathcal{S}_\ell)&&\!=\epsilon_f(N_f^{S_m}\!-\!1)+\epsilon_f(N_f^{S_\ell}\!-\!1)\nonumber\\
&&\!=\epsilon_f(N_f^{S_m}\!+\!N_f^{S_\ell}\!-\!2)\nonumber\\
&&\!<\epsilon_f(N_f^{S_m}\!+\!N_f^{S_\ell}\!-\!1)\!=\!v(\mathcal{S}_m\!\cup\!{\mathcal{S}_\ell}).\label{Equ.3}
\end{eqnarray}
(C2) $N_e^{S_m}\!>\!0$ and $N_e^{S_\ell}\!\!=\!0$:\\
As $N_e^{S_m}\!\geq\!1$, $N_e^{S_m}\!+\!N_e^{S_\ell}\!\geq\!{1}$ holds. In addition, $N_e^{S_{\ell}}\!=\!0$ and $N_f^{S_\ell}\!\geq\!{1}$. These yield
\begin{eqnarray}
v(\mathcal{S}_m)\!+\!v(\mathcal{S}_\ell)&&\!=\epsilon_e(N_e^{S_m}\!-\!1)+\epsilon_fN_f^{S_m}+\epsilon_f(N_f^{S_\ell}\!-\!1)\nonumber\\
&&\!=\epsilon_e(N_e^{S_m}\!-\!1)+\epsilon_f(N_f^{S_m}\!+\!N_f^{S_\ell})-\epsilon_f\nonumber\\
&&\!=v(\mathcal{S}_m\!\cup\!{\mathcal{S}_\ell})-\epsilon_f\!<\!{v(\mathcal{S}_m\!\cup\!{\mathcal{S}_\ell})}.\label{Equ.4}
\end{eqnarray}
The same result goes for $N_e^{S_m}\!=\!0$ and $N_e^{S_\ell}\!>\!0$.

(C3) $N_e^{S_m}\!>\!0$ and $N_e^{S_\ell}\!>\!0$:\\
According to the second case of \eqref{Equ.2}, we derive that
\begin{eqnarray}
v(\mathcal{S}_m)\!+\!v(\mathcal{S}_\ell)&&\!=\epsilon_e(N_e^{S_m}\!+\!N_e^{S_\ell}\!-\!2)\!+\!\epsilon_f(N_f^{S_m}\!+\!N_f^{S_\ell})\nonumber\\
&&\!=v(\mathcal{S}_m\!\cup\!{\mathcal{S}_\ell})\!-\!\epsilon_e\!<\!v(\mathcal{S}_m\!\cup\!{\mathcal{S}_\ell}).\label{Equ.5}
\end{eqnarray}
From \eqref{Equ.3}-\eqref{Equ.5}, the result follows. \hfill $\square$ 

Lemma~\ref{Lemma1} reveals that forming larger coalitions leads to higher total benefit in game $G$. Therefore, the grand coalition $\mathcal{N}$ constitutes the optimal coalition structure with the maximum total platooning benefit. This addresses the question raised in (i). 
\vspace{-3pt}

\begin{remark}\label{Remark2}
    Although forming the largest possible coalition maximizes the collective payoff in game $G$, practical constraints, such as safety requirements or operational limitations, may restrict the size of a coalition, corresponding to the size of a platoon. Nevertheless, our results in Lemma~\ref{Lemma1} show that within the allowable range (i.e., $N\!\leq\!{M}$), forming the largest coalition remains the most beneficial coalition structure for game $G$, yielding the highest collective payoff. For $N\!>\!M$, the grand coalition becomes infeasible, and the trucks must be partitioned into multiple coalitions. We refer readers to our recent work~\cite{Bai2026}, which extends this framework to bounded coalition sizes and develops a stable type-based payoff allocation scheme.
\end{remark}

\subsection{Core-stable payoff allocation}\label{Section3.2}
Next, we propose a payoff allocation scheme that distributes the total platooning benefit $v(\mathcal{N})$ among all participating trucks while ensuring core stability.   
\vspace{-3pt}

Let $x\!=\![x_1,\dots,x_N]^{\top}\!\!\in\!{\mathbb{R}^N}$ with $2\!\leq\!{N}\!\leq\!{M}$, denote a payoff vector of game $G$, where each component $x_i\!\in\!{x}$, $i\!=\!1,\dots,N$, represents the payoff allocated to player $i$. According to the characteristic function defined in \eqref{Equ.2}, the value of the grand coalition $\mathcal{N}$ is given by
\vspace{-5pt}
\begin{eqnarray}
v(\mathcal{N})&&\!=\max_{L\in\mathcal{N}}v(\mathcal{N},L)\nonumber\\
&&\!=\!\begin{cases}
\!\epsilon_e(N_e\!-\!1)+\epsilon_fN_f, & \text{if~$N_e\!\geq\!{1}$,}\\
\!\epsilon_f(N_f\!-\!1), & \text{if~$N_e\!=\!0$.}
\end{cases}\label{Equ.6}
\end{eqnarray}
As defined earlier, $N_f$ and $N_e$ denote the number of FPTs and ETs in $\mathcal{N}$, respectively. The core-stable payoff allocation scheme is then presented below.

\begin{theorem}\label{Theorem1}
Consider the platoon coalitional game $G\!=\!(\mathcal{N}, v)$, where $v$ is defined in \eqref{Equ.2}. Suppose the payoff vector $x\!=\![x_1,\dots,x_N]^{\top}\!\!\in\!{\mathbb{R}^N}$ is constructed as
\vspace{-5pt}
\begin{eqnarray}
x_i\!:=\!\begin{cases}
\!\xi v(\mathcal{N}), & \text{if~~$i\!=\!L\!\in\!\mathcal{N}_e$, $N_e\!\geq\!{1}$}, \\
\!\xi v(\mathcal{N}), & \text{if~~$i\!=\!L\!\in\!\mathcal{N}_f$, $N_e\!=\!{0}$}, \\
\!(1\!-\!\xi)\epsilon_e, & \text{if~~$i\!=\!F\!\in\!\mathcal{N}_e$}, \\
\!(1\!-\!\xi)\epsilon_f, & \text{if~~$i\!=\!F\!\in\!\mathcal{N}_f$},
\end{cases}\label{Equ.7}
\end{eqnarray}
where $v(\mathcal{N})$ is of the form \eqref{Equ.6}, with $L$ and $F$ denoting the leader and follower of the platoon, and the parameter $\xi\!\in\!{\mathbb{R}_{+}}$ satisfies
\begin{eqnarray}
\begin{cases}
\!0\!<\xi\leq{\frac{\epsilon_e}{\epsilon_e(N_e-1)+\epsilon_fN_f}}, & \text{if~$N_e\!\geq\!{1}$},\\
\!0\!<\xi\leq{\frac{1}{N-1}},& \text{if~$N_e\!=\!0$.} 
\label{Equ.8}
\end{cases}
\end{eqnarray}
Then, the payoff vector $x$ lies in the core of game $G$.
\end{theorem}

\noindent \textbf{Proof.} The proof proceeds in two steps. First, we show that the payoff vector $x$ defined in \eqref{Equ.7} is a valid payoff for game $G$, satisfying conditions (2a) and (2b) in Definition~\ref{Def2}. In the second step, we demonstrate that $x\!\in\!{\mathrm{Core}(G)}$ by showing that it satisfies both the efficiency and coalitional rationality conditions specified in Definition~\ref{Def3}.

(S1) Recall that $N\!\geq\!{2}$ and $\epsilon_e\!<\!\epsilon_f$, thus, $0\!<\!\xi\!\leq\!1$ following \eqref{Equ.8}. By \eqref{Equ.6} and definition \eqref{Equ.7}, $x_i\!>\!0$. As $v(\{i\})\!=\!0$ by \eqref{Equ.2}, we have  
\begin{eqnarray}
x_i\!\geq\!{v(\{i\})},~~\forall{i}\!\in\!\mathcal{N}.\nonumber
\end{eqnarray}
This meets condition (2a) in Definition~\ref{Def2}. As revealed by Lemma~\ref{Lemma1}, the optimal coalition structure is $\mathcal{P}\!=\!\{\mathcal{N}\}$, Below, we will show that $\sum_{i\in\mathcal{N}}\!x_i\!=\!v(\mathcal{N})$ holds true. 

If $N_e\!\geq\!{1}$, by \eqref{Equ.6} and \eqref{Equ.7}, we can obtain that
\begin{eqnarray}
\sum_{i\in{\mathcal{N}}}x_i\!&&=x_L+\!\!\sum_{i\in{\mathcal{N}_e\setminus{\{L\}}}}\!\!\!\!\!x_i+\!\sum_{i\in{\mathcal{N}_f}}\!x_i\nonumber\\
&&=\xi{v(\mathcal{N})}+(1\!-\!\xi)\big(\epsilon_e(N_e\!-\!1)\!+\!\epsilon_fN_f\big)\nonumber\\
&&=\xi{v(\mathcal{N})}+(1\!-\!\xi)v(\mathcal{N})\nonumber\\
&&=v(\mathcal{N}).\nonumber
\end{eqnarray}
If $N_e\!=\!0$, i.e., the leader of the platoon is selected from the FPTs and $N\!=\!N_f$. In this case, we have
\begin{eqnarray}
\sum_{i\in{\mathcal{N}}}x_i\!&&=x_L+\!\!\sum_{i\in{\mathcal{N}\setminus\{L\}}}\!\!\!\!\!x_i\nonumber\\
&&=\xi{v(\mathcal{N})}+(1\!-\!\xi)\epsilon_f(N_f\!-\!1)\nonumber\\
&&=\xi{v(\mathcal{N})}+(1\!-\!\xi)v(\mathcal{N})=v(\mathcal{N}).\nonumber
\end{eqnarray}
Thus, condition (2b) in Definition~\ref{Def2} is satisfied. This proves that $x$ is a valid payoff allocation for $G$.
\vspace{-5pt}

(S2) Next, we demonstrate that $x\!\in\!{\mathrm{Core}(G)}$. As established above, the efficiency condition in Definition~\ref{Def3} is already satisfied. Below, we prove the coalitional rationality, i.e., $\sum_{i\in{\mathcal{S}}}x_i\!\geq\!{v(\mathcal{S})},\forall{\mathcal{S}}\!\subseteq\!{\mathcal{N}}$. Since the leader of the grand coalition $\mathcal{N}$ may not be included in coalition $\mathcal{S}$, we consider the following two cases separately.
\vspace{-5pt}

(C1) The leader of $\mathcal{N}$ is included in $\mathcal{S}$: \\ Recall $v(\mathcal{S})$ defined in \eqref{Equ.2}. For any $\mathcal{S}\subseteq{\mathcal{N}}$, according to the values of $N^S$ and $N_e^S$, and following \eqref{Equ.7}, we have
\vspace{-5pt}
\begin{eqnarray}
&&\!\sum_{i\in\mathcal{S}}^{}x_i=\nonumber\\
&&\!\begin{cases}
\!0 & \!\!\!\!\text{if~$N^S\!\!=\!0$,}\\
\!\xi{v(\mathcal{N})}\!+\!(1\!-\!\xi)\big(\epsilon_e(N_e^S\!\!-\!1)\!+\!\epsilon_fN_f^S\big),&\!\!\!\!\text{if $N^S\!\!\geq\!{1}$, $N_e^S\!\!\geq\!{1}$,}\\
& \  \text{$L\!\in\!\mathcal{N}_e^{\mathcal{S}}$,}\\
\!\xi v(\mathcal{N})\!+\!(1\!-\!\xi)\epsilon_f(N_f^S\!\!-\!1), & \!\!\! \text{if $N^S\!\!\geq\!{1}$,$N_e^S\!\!=\!{0}$,}\\
& \ \text{$L\!\in\!\mathcal{N}_f^{\mathcal{S}}$.}
\end{cases}\nonumber
\end{eqnarray}
For $N^S\!=\!0$, there is $\sum_{i\in\mathcal{S}}x_i\!=\!v(\mathcal{S})\!=\!0$. By Lemma~\ref{Lemma1}, $G$ is superadditive. Thus, the grand coalition maximizes $v$, i.e., $v(\mathcal{N})\!\geq\!{v(\mathcal{S})}$, $\forall\mathcal{S}\!\!\subseteq\!{\mathcal{N}}$. For $N^S\!\geq\!{1}$, integrating \eqref{Equ.2} into the above result, we obtain
\begin{eqnarray}
\sum_{i\in\mathcal{S}}^{}x_i\!&&=
\xi{v(\mathcal{N})}+(1\!-\!\xi)v(\mathcal{S})\nonumber\\
&&\geq{\xi{v(\mathcal{S})}+(1\!-\!\xi)v(\mathcal{S})}\!=\!v(\mathcal{S}).\label{Equ.9}
\end{eqnarray}

(C2) The leader of $\mathcal{N}$ is not included in $\mathcal{S}$:\\ In this case, all the trucks in $\mathcal{S}$ are followers in $\mathcal{N}$. In line with the payoff allocation defined in~\eqref{Equ.7}, we derive that
\begin{eqnarray}
\sum_{i\in{\mathcal{S}}}x_i\!=(1\!-\!\xi)(\epsilon_eN_e^S\!+\!\epsilon_fN_f^S).\label{Equ.10}
\end{eqnarray}
If $N^S\!=\!0$, i.e., $N_e^S\!=\!N_f^S\!=\!0$, then $\sum_{i\in{\mathcal{S}}}\!x_i\!=\!0\!=\!v(\mathcal{S})$. If $N^S\!\geq\!{1}$, since $\epsilon_e\!<\!\epsilon_f$, and depending on the value of $N_e^S$, we have
\begin{eqnarray}
v(\mathcal{S})&&\!\leq\max\big\{\epsilon_e(N_e^S\!\!-\!1)\!+\!\epsilon_fN_f^S,\epsilon_f(N_f^S\!\!-\!1)\big\}\nonumber\\
&&\!=\epsilon_eN_e^S\!+\epsilon_fN_f^S\!-\epsilon_e.\label{Equ.11}
\end{eqnarray}
As the leader of $\mathcal{N}$ is not included in $\mathcal{S}$, we have $N_f^S\!+\!N_e^S\!\leq\!{N_f\!+\!N_e\!-\!1}$, which is equivalent to
\begin{eqnarray}
(N_e\!-\!N_e^S)+(N_f\!-\!N_f^S)\!\geq\!{1}.\nonumber
\end{eqnarray}
This leads to the following inequality: 
\begin{eqnarray}
&&\epsilon_e(N_e\!-\!N_e^S)+\epsilon_f(N_f\!-\!N_f^S)\nonumber\\
&&>{\epsilon_e}(N_e\!-\!N_e^S)+\epsilon_e(N_f\!-\!N_f^S)\nonumber\\
&&={\epsilon_e}\big((N_e\!-\!N_e^S)+(N_f\!-\!N_f^S)\big)\!\geq\!{\epsilon_e}.\label{Equ.12}
\end{eqnarray}
Rewriting \eqref{Equ.12} gives us 
\begin{eqnarray}
\epsilon_eN_e^S+\epsilon_fN_f^S<{\epsilon_e(N_e\!-\!1)+\epsilon_fN_f}.\label{Equ.13}
\end{eqnarray}
According to \eqref{Equ.8}, if $N_e\!\geq\!{1}$, there is $0\!<\!\xi\!\leq\!{\frac{\epsilon_e}{\epsilon_e(N_e-1)+\epsilon_fN_f}}$. Then, combining \eqref{Equ.10}, \eqref{Equ.11} and \eqref{Equ.13} yields 
\begin{eqnarray}
&&v(\mathcal{S})-\!\sum_{i\in\mathcal{S}}x_i\nonumber\\
&&\leq\epsilon_eN_e^S+\epsilon_fN_f^S\!-\epsilon_e-(1\!-\!\xi)(\epsilon_eN_e^S\!+\!\epsilon_fN_f^S)\nonumber\\
&&=\xi(\epsilon_eN_e^S\!+\!\epsilon_fN_f^S)-\epsilon_e\nonumber\\
&&\leq\epsilon_e\!\left(\!\frac{\epsilon_eN_e^S\!+\!\epsilon_fN_f^S}{\epsilon_e(N_e\!-\!1)+\epsilon_fN_f}\!-\!1\!\right)<{0}.\label{Equ.14}
\end{eqnarray}
In the other case where $N_e\!=\!0$, there is $N_f\!=\!N\!\geq\!{2}$. Consequently, $N_e^S\!=\!0$ and $N_f^S\!\leq\!{N\!-\!1}$. From~\eqref{Equ.8}, it follows that $0\!<\!\xi\!\leq\!{\frac{1}{N-1}}$. Since $v(\mathcal{S})\!=\!\epsilon_f(N_f^S\!-\!1)$ by \eqref{Equ.2}, and consistent with \eqref{Equ.10}, we derive that
\begin{eqnarray}
v(\mathcal{S})\!-\!\sum_{i\in\mathcal{S}}x_i&&\!=\epsilon_f(N_f^S\!-\!1)-(1\!-\!\xi)\epsilon_fN_f^S\nonumber\\
&&\!=\epsilon_f(\xi N_f^S\!-\!1)\!\leq\!{0}.\label{Equ.15}
\end{eqnarray}
With \eqref{Equ.9}, \eqref{Equ.14} and \eqref{Equ.15}, we demonstrate that $\sum_{i\in{\mathcal{S}}}^{}\!x_i\!\geq\!{v(\mathcal{S})}$ holds for any $\mathcal{S}\!\subseteq\!{\mathcal{N}}$. 

Given the above results, we prove that the payoff vector $x$ designed in \eqref{Equ.7} with $\xi$ satisfying \eqref{Equ.8} lies in the core of the game $G$. This completes the proof of Theorem~\ref{Theorem1}. \hfill $\square$ 
\vspace{-5pt}

Theorem~\ref{Theorem1} presents a payoff allocation scheme to design the payoff vector $x$ for the platoon coalitional game $G$, which distributes the total platooning benefit $v(\mathcal{N})$ among all participants in a stable manner, ensuring that $x\!\in\!\mathrm{Core}(G)$. For a general mixed-energy truck platoon with $N_e\!\geq\!{1}$, the payoff $x_i$ defined in~\eqref{Equ.7} depends implicitly on both $\epsilon_e$ and $\epsilon_f$ through $\xi$ and $v(\mathcal{N})$. The resulting allocation guarantees that no truck has an incentive to deviate from the grand coalition $\mathcal{N}$ in pursuit of higher platooning benefits by forming alternative coalitions.
\vspace{-3pt}

\begin{remark}
In contrast to existing parameterized or compromise-type allocation approaches in cooperative game theory, e.g.,~\cite{tijs1993compromise}; \cite{sun2023general}, Theorem~\ref{Theorem1} establishes a problem-specific analytical framework for benefit allocation in mixed-energy truck platooning. In particular, the parameter $\xi$ is introduced to capture the payoff imbalance arising from both the leader-follower asymmetry and the heterogeneous follower savings across truck types. In addition, Theorem~\ref{Theorem1} provides a closed-form payoff allocation for each truck with an explicit admissible range of $\xi$ under which the resulting allocation is ensured to lie in the core.
\end{remark}

\begin{corollary}\label{Corollary1}
As a special case of Theorem~\ref{Theorem1}, if the trucks in~$\mathcal{N}$ are homogeneous (i.e., either all FPTs or ETs), and $\xi$ in \eqref{Equ.7} satisfies
\begin{eqnarray}
0<\xi\leq{\frac{1}{N\!-\!1}},\label{Equ.16}
\end{eqnarray}
then, the payoff vector $x$ defined in \eqref{Equ.7} is core-stable. 
\end{corollary}

\begin{remark}\label{Remark3} 
The second condition of \eqref{Equ.8} captures the special case where all trucks in $\mathcal{N}$ are fuel-powered, while the first condition reduces to~\eqref{Equ.16} when $N_f\!=\!0$. It is important to note that the even-split payoff allocation mechanism, as widely adopted in~\citep{9435129,10209062}, inherently satisfies the core stability condition given in \eqref{Equ.16}. This scheme allocates an equal share of the total value, $\frac{1}{N}v(\mathcal{N})$, to each truck, which corresponds to setting $\xi\!=\!\frac{1}{N}$ in the proposed payoff structure defined in \eqref{Equ.7}. By Corollary~\ref{Corollary1}, this results in a payoff vector that lies within the core of game $G$. 
\end{remark}
\vspace{-5pt}

\section{Stable and fair benefit allocation}\label{Section 4}
The allocation scheme proposed in Section~\ref{Section 3} is core-stable and ensures that the grand coalition remains intact. However, it may not fully reflect the individual contributions of each truck to the overall platooning benefit. A stable payoff allocation yet perceived as unfair can leave truck owners feeling undervalued, even though they have no better alternatives. Enhancing fairness within a stable allocation helps strengthen platoon cohesion, mitigates dissatisfaction, and reduces the incentive to defect, thereby promoting more sustainable and enduring cooperation among trucks.    

This section presents a payoff allocation scheme that achieves both core stability and fairness. We first derive a closed-form Shapley value-based allocation that distributes the total platooning benefit equitably and efficiently among participating trucks. Subsequently, we establish sufficient conditions under which the Shapley value-based allocation ensures core stability.       

\subsection{Shapley value-based allocation}\label{Section4.1}
In coalitional game theory, the Shapley value provides a principled approach to fair payoff allocation by assigning each player their expected marginal contribution averaged over all possible coalitions (or orderings); see~\cite{roth1988shapley,reddy2013time}. While the nucleolus represents an alternative fairness-oriented solution concept, it is typically employed when the core is empty or excessively large. As such conditions do not arise in the platoon coalitional game $G$, the Shapley value is adopted to enhance fairness. Next, we introduce the formal definition of the Shapley value.

\begin{definition} \cite[p.~126]{peters2008game}
\label{Def5}
Consider a coalitional game $G\!=\!(\mathcal{N},v)$ with transferable utility as defined in Definition~\ref{Def1}. The Shapley value is a payoff allocation $\phi\!=\![\phi_1,\dots,\phi_N]^{\top}\!\!\in\!\mathbb{R}^N$, where each player $i\!\in\!\mathcal{N}$ receives  
\begin{eqnarray}
\phi_i=\!\!\!\sum_{\mathcal{S}\subseteq\mathcal{N}\setminus\{i\}}\!\!\!\!\!\frac{|\mathcal{S}|!(N\!-\!|\mathcal{S}|\!-\!1)!}{N!}\big(v(\mathcal{S}\!\cup\!\{i\})\!-\!v(\mathcal{S})\big).\label{Equ.17}
\end{eqnarray}
\end{definition}
\vspace{-10pt}
In what follows, we present the Shapley value-based payoff allocation for the platoon coalitional game $G$.
\vspace{-5pt}

\begin{proposition}\label{Proposition1}
Consider the platoon coalitional game $G\!=\!(\mathcal{N},v)$ where $v$ is defined in \eqref{Equ.2}. The closed-form Shapley value-based payoff for truck $i\!\in\!\mathcal{N}$ is given by
\begin{eqnarray}
\phi_i=\!\begin{cases}
\!\!(1\!-\!\frac{1}{N_e})\epsilon_e\!+\!\frac{N_f}{NN_e}\epsilon_f, & \text{if $i\!\in\!\mathcal{N}_e$,}\\
\!\!(1\!-\!\frac{1}{N})\epsilon_f,&\text{if~$i\!\in\!\mathcal{N}_f$}.
\end{cases}\label{Equ.18}
\end{eqnarray}
\end{proposition}
\vspace{-12pt}
\noindent \textbf{Proof.} The proof consists of three steps. First, we derive the Shapley values for each of ET and FPT. Then, we verify that the resulting allocation satisfies the efficiency condition, i.e., $\sum_{i\in\mathcal{N}}\phi_i\!=\!v(\mathcal{N})$. By \eqref{Equ.2}, the value of a coalition $\mathcal{S}\!\subseteq\!{\mathcal{N}}$ depends not only on its composition but also on the ordering of trucks in $\mathcal{S}$, which determines the probability of each truck being selected as the leader to maximize $v(\mathcal{S})$. Below, to compute the Shapley value for any truck $i\!\in\!\mathcal{N}$ by~\eqref{Equ.17}, we focus on coalitions $\mathcal{S}\!\subseteq\!{\mathcal{N}\setminus}\{i\}$ for which the marginal contribution of truck $i$ is nonzero.

(S1) Shapley value of each ET ($i\!\in\!\mathcal{N}_e$ with $N_e\!\neq\!{0}$):
\vspace{-5pt}

(C1-1) $\mathcal{S}\!\subseteq\!{\mathcal{N}\!\setminus\!\{i\}}$ includes no ETs but at least one FPT:\\
By~\eqref{Equ.2}, we have $v(\mathcal{S}\!\cup\!\{i\})\!-\!v(\mathcal{S})\!=\!\epsilon_f$. The probability that $\mathcal{S}$ satisfying the condition (C1-1) is
\begin{eqnarray}
\Pr(\text{C1-1})\!&&=\frac{\sum_{k=1}^{N_f}\!\binom{N_f}{k}\cdot{k!}\cdot(N\!-\!1\!-\!k)!}{N!}\nonumber\\
&&=\frac{1}{N!}\!\left(\sum_{k=1}^{N_f}\frac{N_f!}{(N_f\!-\!k)!}\cdot(N\!-\!1\!-\!k)!\right)\nonumber\\
&&=\frac{N_f!}{N!}\!\left(\sum_{p=0}^{N_f-1}\frac{\big(N\!-\!1\!-\!(N_f\!-\!p)\big)!}{p!}\right)\nonumber\\
&&=\frac{N_f!}{N!}\!\left(\sum_{p=0}^{N_f-1}\frac{(N_e\!+p\!-\!1)!}{p!}\right)\nonumber\\
&&=\frac{N_f!(N_e\!-\!1)!}{N!}\!\left(\sum_{p=0}^{N_f-1}\!\binom{N_e\!+\!p\!-\!1}{p}\!\right)\nonumber\\
&&=\frac{N_f!(N_e\!-\!1)!}{N!}\binom{N_e\!+\!N_f\!-\!1}{N_f\!-\!1}\nonumber\\
&&=\frac{N_f!(N_e\!-\!1)!}{N!}\cdot\frac{(N_e\!+\!N_f\!-\!1)!}{(N_f\!-\!1)!N_e!}\nonumber\\
&&=\frac{N_f(N\!-\!1)!}{N!N_e}=\frac{N_f}{NN_e},\label{Equ.19}
\end{eqnarray}
where $p\!=\!N_f\!-\!k$ and $N\!=\!N_f\!+\!N_e$. Note that the binomial sum identity $\sum_{q=0}^{n}\binom{r+q}{q}\!=\!\binom{r+n+1}{n}$, where $r,n\!\in\!\mathbb{N}$, is used in the derivation; see~\cite{graham1994concrete}. 
\vspace{-5pt}

(C1-2) $\mathcal{S}\!\subseteq\!{\mathcal{N}\!\setminus\!\{i\}}$ includes at least one ET:\\
In this case, the marginal contribution of truck $i$ is $v(\mathcal{S}\!\cup\!\{i\})\!-\!v(\mathcal{S})\!=\!\epsilon_e$. The probability that at least one ET precedes truck $i$ equals one minus the probability that all preceding trucks are FPTs. By~\eqref{Equ.19}, we obtain
\begin{eqnarray}
\Pr(\text{C1-2})\!&&=1-\frac{\sum_{k=0}^{N_f}\binom{N_f}{k}\cdot{k!}\cdot(N\!-\!1\!-\!k)!}{N!}\nonumber\\
&&=1-\left(\frac{(N\!-\!1)!}{N!}\!+\!\frac{N_f}{NN_e}\right)\nonumber\\
&&=1-\left(\frac{1}{N}\!+\!\frac{N\!-\!N_e}{NN_e}\right)=1-\frac{1}{N_e}.\nonumber
\end{eqnarray}
Combining the two cases above, the Shapley value of each ET $i\!\in\!{\mathcal{N}_e}$ takes the form of
\begin{eqnarray}
\phi_i&&\!\!=\!\Pr(\text{C1-1})\epsilon_f\!+\!\Pr(\text{C1-2})\epsilon_e\!=\!\frac{N_f}{NN_e}\epsilon_f\!+\!(1\!-\!\frac{1}{N_e})\epsilon_e.\label{Equ.20}
\end{eqnarray}
\vspace{-18pt}

(S2) Shapley value of each FPT ($i\!\in\!\mathcal{N}_f$ with $N_f\!\neq\!{0}$):
\vspace{-5pt}

(C2-1) $\mathcal{S}\!\subseteq\!{\mathcal{N}\!\setminus\!\{i\}}$ includes at least one FPT:\\
Despite the number of ETs in coalition $\mathcal{S}$, if $\mathcal{S}$ includes at least one FPT, then any truck $i\!\in\!\mathcal{N}_f$ joining the coalition contributes the benefit of one FPT follower. Thus, the marginal contribution of truck $i$ is $v(\mathcal{S}\!\cup\!\{i\})\!-\!v(\mathcal{S})\!=\!\epsilon_f$.

Accordingly, the probability that condition (C2-1) holds for coalition $\mathcal{S}$ is
\begin{eqnarray}
\Pr(\text{C2-1})=1-\frac{(N\!-\!1)!}{N!}=1-\frac{1}{N}.\nonumber
\end{eqnarray}
This is derived by excluding the case where truck $i$ is the first in the ordering. Thus, the Shapley value of truck $i\!\in\!\mathcal{N}_f$ is denoted as
\begin{eqnarray}
\phi_i=\Pr(\text{C2-1})\epsilon_f=(1\!-\!\frac{1}{N})\epsilon_f.\label{Equ.21}
\end{eqnarray}
(S3) Lastly, the efficiency of the resulting payoff is verified. By~\eqref{Equ.2}, \eqref{Equ.20} and \eqref{Equ.21}, if $N_e\!\geq\!{1}$, we have
\begin{eqnarray}
\sum_{i\in\mathcal{N}}\phi_i&&=\sum_{i\in\mathcal{N}_e}\phi_i+\sum_{i\in\mathcal{N}_f}\phi_i\nonumber\\
&&=\left(\frac{N_f}{NN_e}\epsilon_f+(1\!-\!\frac{1}{N_e})\epsilon_e\right)\!N_e+(1\!-\!\frac{1}{N})\epsilon_fN_f\nonumber\\
&&=\frac{N_f}{N}\epsilon_f+(N_e\!-\!1)\epsilon_e+N_f\epsilon_f-\frac{N_f}{N}\epsilon_f\nonumber\\
&&=(N_e\!-\!1)\epsilon_e\!+\!N_f\epsilon_f=v(\mathcal{N}).\label{Equ.22}
\end{eqnarray}
If $N_e\!=\!0$, then $N_f\!=\!N$. In this case, it follows  
\begin{eqnarray}
\sum_{i\in\mathcal{N}}\phi_i&&=\!\sum_{i\in\mathcal{N}_f}\phi_i\nonumber\\
&&=(1\!-\!\frac{1}{N})\epsilon_fN_f=(N\!-\!1)\epsilon_f=v(\mathcal{N}).\label{Equ.23}
\end{eqnarray}
Given that \eqref{Equ.20}-\eqref{Equ.23} hold, we complete the proof of Proposition~\ref{Proposition1}.  \hfill $\square$ 

Proposition~\ref{Proposition1} characterizes the payoff for each truck in $\mathcal{N}$ as proportional to its expected marginal contribution, achieving a fair allocation of the total platooning benefits among all participating trucks. The results show that the Shapley value-based payoff of each ET is comprised of two components: (i) the expected payoff from joining platoons as a follower when other ETs lead, denoted as $(1\!-\!\frac{1}{N_e})\epsilon_e$; and (ii) the expected payoff from enabling FPT followers when acting as the leader, captured by $\frac{N_f}{NN_e}\epsilon_f$. This indicates that an increase in the number of ETs in $\mathcal{N}$ reduces each ET's chance to lead, thereby leading to lower leadership gains, whereas it increases its expected benefit as a follower. In contrast, each FPT's payoff is determined by the platoon size $N$, characterized by $(1\!-\!\frac{1}{N})\epsilon_f$. The larger the platoon, the greater the payoff each FPT can receive. The structural distinction between the Shapley values of ETs and FPTs, as given in~\eqref{Equ.18}, also indicates that ETs tend to join mixed-energy platoons with a high proportion of FPTs, whereas FPTs prioritize forming larger platoons in which their marginal contributions remain significant.

\begin{remark}\label{Remark5}
Computing Shapley values involves evaluating each player's marginal contribution across all possible coalitions, resulting in combinatorial explosions and high computational complexity. Our approach addresses this challenge by proposing a closed-form solution for the Shapley value-based payoff allocation. It eliminates the need for exhaustive enumeration and ensures scalability and computational efficiency, making the proposed scheme well-suited for practical implementation, even in systems with large platoons. 
\end{remark}

The following result establishes the connection between the stable payoff allocation proposed in Theorem~\ref{Theorem1} and the fair, Shapley value–based allocation presented in Proposition~\ref{Proposition1}.

\begin{proposition}\label{Proposition2}
Consider two special cases of Proposition~\ref{Proposition1}: (a) all trucks in $\mathcal{N}$ are homogeneous (i.e., either all FPTs or all ETs); and (b) the trucks are heterogeneous, but only one ET exists in $\mathcal{N}$ (i.e., $N_e\!=\!1$). In both cases, the following results hold:\\
(i) The Shapley value-based payoff $\phi_i$ determined by \eqref{Equ.18} coincides with the payoff $x_i$ defined in \eqref{Equ.7} for any $i\!\in\!\mathcal{N}$, with the special choice $\xi\!=\!\frac{1}{N}$;\\
(ii) The payoff vector $\phi\!=\![\phi_1,\dots,\phi_N]^{\top}\!\!\in\!\mathbb{R}^N$ given by~\eqref{Equ.18} lies in the core of game $G$, i.e., $\phi\in\mathrm{Core}(G)$. 
    
\end{proposition}
\noindent \textbf{Proof.} (i-a) By \eqref{Equ.18}, if all trucks in $\mathcal{N}$ are homogeneous, one has
\begin{eqnarray}
\phi_i\!=\!\begin{cases}
\!\!(1\!-\!\frac{1}{N})\epsilon_e, &\text{if~$i\!\in\!\mathcal{N}_e\!=\!\mathcal{N}$},\\
\!\!(1\!-\!\frac{1}{N})\epsilon_f, &\text{if~$i\!\in\!\mathcal{N}_f\!=\!\mathcal{N}$}.\label{Equ.24}
\end{cases}
\end{eqnarray}
Let us consider the first case where $\mathcal{N}_e\!=\!\mathcal{N}$. Suppose that $\xi\!=\!\frac{1}{N}$. Following the payoff defined in \eqref{Equ.7}, if $i\!=\!L\!\in\!\mathcal{N}_e$, the payoff allocated to the leader is
\begin{eqnarray}
x_L=\xi{v(\mathcal{N})}=\xi{(N\!-\!1)\epsilon_e}=(1\!-\!\frac{1}{N})\epsilon_e.\nonumber
\end{eqnarray}
If $i\!=\!F\!\in\!\mathcal{N}_e$, by \eqref{Equ.7}, we have 
\begin{eqnarray}
x_F\!=\!(1\!-\!\xi)\epsilon_e=(1\!-\!\frac{1}{N})\epsilon_e.\nonumber
\end{eqnarray}
Both $x_L$ and $x_F$ align with the Shapley value-based payoff given in \eqref{Equ.24}. The case where $\mathcal{N}_f\!=\!\mathcal{N}$ can be proved similarly. 

(i-b) If trucks in $\mathcal{N}$ are heterogeneous while satisfying $N_e\!=\!1$, then $N_f\!=\!N\!-\!1$. By \eqref{Equ.18}, the Shapley value of each ET $i\!\in\!\mathcal{N}_e$ is 
\begin{eqnarray}
\phi_i=\frac{N\!-\!1}{N}\epsilon_f=(1\!-\!\frac{1}{N})\epsilon_f.\nonumber
\end{eqnarray}
Accordingly, the Shapley value of each FPT $i\!\in\!\mathcal{N}_f$ is of the same form by \eqref{Equ.18}. As the only one ET serves as the leader in the platoon by \eqref{Equ.2}, in line with \eqref{Equ.7}, the payoff of truck $i\!=\!L\!\in\!\mathcal{N}_e$ is  
\begin{eqnarray}
x_L\!=\!\xi{v(\mathcal{N})}=\xi\epsilon_fN_f=(1\!-\!\frac{1}{N})\epsilon_f.\nonumber
\end{eqnarray}
Since $\xi\!=\!\frac{1}{N}$, the payoff of each follower $i\!=\!F\!\in\!\mathcal{N}_f$ follows
\begin{eqnarray}
x_F\!=\!(1\!-\!\xi)\epsilon_f\!=\!(1\!-\!\frac{1}{N})\epsilon_f.\nonumber
\end{eqnarray}
This proves that $\phi_i\!=\!x_i$ for any $i\!\in\!\mathcal{N}$, with $\xi\!=\!\frac{1}{N}$ in the two cases of Proposition~\ref{Proposition2}.

(ii) In case (a), as $0\!<\!\xi\!=\!\frac{1}{N}\!<\!\frac{1}{N-1}$, by Corollary~\ref{Corollary1}, $\phi\!\in\!{\mathrm{Core}(G)}$ holds. In case (b), $N_e\!=\!1$. Since $\phi_i\!=\!(1\!-\!\frac{1}{N})\epsilon_f$, $\forall{i\in\mathcal{N}}$ by \eqref{Equ.18}, every truck receives the same payoff. The efficiency of $\phi$ follows from Proposition~\ref{Proposition1}. For any coalition $\mathcal{S}\!\subseteq\!\mathcal{N}$, if $N_e^S\!=\!0$, then $v(\mathcal{S})\!=\!\epsilon_f(N_f^S\!-\!1)$ and 
\begin{eqnarray}
\sum_{i\in\mathcal{S}}\phi_i-v(\mathcal{S})=
\epsilon_f\Big(1-\frac{N_f^{S}}{N}
\Big)\!\ge\!0.\nonumber
\end{eqnarray}
If $N_e^S\!=\!1$, we have $v(\mathcal{S})\!=\!\epsilon_f N_f^S$, and 
\begin{eqnarray}
\sum_{i\in\mathcal{S}}\phi_i-v(\mathcal{S})=
\epsilon_f\Bigl(\frac{N\!-\!1\!-\!N_f^S}{N}\Bigr)\!\ge\! 0.\nonumber
\end{eqnarray}
Thus, $\sum_{i\in\mathcal{S}}\!\phi_i\!\ge\! v(\mathcal{S})$ for all $\mathcal{S}\!\subseteq\!\mathcal{N}$, and therefore $\phi\!\in\!{\mathrm{Core}(G)}$. \hfill $\square$ 

Proposition~\ref{Proposition2} reveals that when the trucks in $\mathcal{N}$ are homogeneous or there is only one ET, the Shapley value-based payoff reduces to the same form as the payoff allocation proposed in Theorem~\ref{Theorem1}, with $\xi\!=\!\frac{1}{N}$. Moreover, the Shapley value lies in the core of game $G$, indicating that the even-split allocation is both core-stable and fair. 

\subsection{Core stability of the Shapley value}\label{Section4.2}
This subsection identifies the conditions under which the Shapley value-based payoff allocation proposed in Proposition~\ref{Proposition1} lies within the core of game $G$, thereby ensuring both stability and fairness. We start by presenting the necessary and sufficient condition for core stability, as stated in Lemma~\ref{Lemma2}.

\begin{lemma}\label{Lemma2}
The Shapley value-based payoff allocation $\phi\!=\![\phi_1,\dots,\phi_N]^{\top}\!\!\in\!\mathbb{R}$ presented in Proposition~\ref{Proposition1} with $N_f\!>\!0$ and $N_e\!>\!0$ lies in the core of game $G\!=\!(\mathcal{N},v)$, if and only if, for any coalition $\mathcal{S}\!\subseteq\!{\mathcal{N}}$, the following holds: 
\begin{eqnarray}
\frac{\epsilon_e}{\epsilon_f}\geq{\frac{1}{N}\cdot\frac{N_f^SN_e\!-\!N_fN_e^S}{N_e\!-\!N_e^S}}, \quad\text{for~~} 1\!\leq\!N_e^S\!<\!N_e,\label{Equ.25}
\end{eqnarray}
 where $N_e^S\!\!=\!|\mathcal{S}\!\cap\!{\mathcal{N}_e}|$ and $N_f^S\!\!=\!|\mathcal{S}\!\cap\!{\mathcal{N}_f}|$.
\end{lemma}
\noindent \textbf{Proof.} The proof proceeds in two steps, i.e., necessity and sufficiency, as detailed below.

(S1) Necessity ($\phi\!\in\!{\mathrm{Core}(G)}\Rightarrow$ Inequality \eqref{Equ.25} holds):\\
(C1) If $1\!\leq\!N_e^S\!<\!N_e$: Given that $\phi\!\in\!\mathrm{Core}(G)$, by Definition~\ref{Def3}, $\forall{\mathcal{S}\subseteq{\mathcal{N}}}$, there is $\sum_{i\in\mathcal{S}}\!\phi_i\!\geq\!{v(\mathcal{S})}$. By \eqref{Equ.18} and~\eqref{Equ.2}, we have
\begin{eqnarray}
\sum_{i\in\mathcal{S}}\phi_i&&\!=\!\big((1\!-\!\frac{1}{N_e})\epsilon_e\!+\!\frac{N_f}{NN_e}\epsilon_f\big)N_e^S+(1\!-\!\frac{1}{N})\epsilon_fN_f^S\nonumber\\
&&\!\geq{v(\mathcal{S})}\!=\!(N_e^S\!-\!1)\epsilon_e\!+\!\epsilon_fN_f^S,\nonumber
\end{eqnarray}
which is simplified as
\begin{eqnarray}
\epsilon_e(1\!-\!\frac{N_e^S}{N_e})\geq{\epsilon_f(\frac{N_f^S}{N}\!-\!\frac{N_fN_e^S}{NN_e})}.\label{Equ.26}
\end{eqnarray}
As $\epsilon_f\!>\!0$, if ${N_e^S}\!<\!{N_e}$, \eqref{Equ.26} can be rewritten as
\begin{eqnarray}
\frac{\epsilon_e}{\epsilon_f}\geq{\frac{1}{N}\!\cdot\!\frac{N_f^SN_e\!-\!N_fN_e^S}{N_e\!-\!N_e^S}}.\nonumber
\end{eqnarray}
(C2) If $N_e^S\!=\!N_e$: Since $N_f^S\leq{N_f}$, by \eqref{Equ.26}, we get
\begin{eqnarray}
0\geq{\frac{\epsilon_f}{N}(N_f^S\!-\!N_f)},\nonumber
\end{eqnarray}
which is always satisfied.
(C3) If $N_e^S\!=\!0$: In this case, by \eqref{Equ.18}, one has
\begin{eqnarray}
\sum_{i\in\mathcal{S}}\phi_i\!=\!(1\!-\!\frac{1}{N})\epsilon_fN_f^S\!\geq\!{v(\mathcal{S})}\!=\!{\epsilon_f(N_f^S\!-\!1)}.\label{Equ.27}
\end{eqnarray}
Since $N_f^S\!\leq\!{N}$, \eqref{Equ.27} is satisfied.

(S2) Sufficiency (Inequality \eqref{Equ.25} holds $\Rightarrow$ $\phi\!\in\!{\mathrm{core}(G)}$):\\
The efficiency of $\phi$ has been proved in (S3) of the proof of Proposition~\ref{Proposition1}. Below, the coalitional rationality in Definition~\ref{Def3}, i.e., $\sum_{i\in\mathcal{S}}\!\phi_i\!\geq\!{v(\mathcal{S})}$, $\forall{\mathcal{S}\!\subseteq\!{\mathcal{N}}}$, is proved.
 
Given that \eqref{Equ.25} holds. For any $\mathcal{S}\!\subseteq\!{\mathcal{N}}$, if $1\!\leq\!{N_e^S}\!<\!{N_e}$, by \eqref{Equ.18} and \eqref{Equ.2}, we derive that
\begin{eqnarray}
&&\sum_{i\in\mathcal{S}}\phi_i\!-v(\mathcal{S})\nonumber\\
&&=\big((1\!-\!\frac{1}{N_e})\epsilon_e\!+\!\frac{N_f}{NN_e}\epsilon_f\big)N_e^S\!+\!(1\!-\!\frac{1}{N})\epsilon_fN_f^S\nonumber\\
&&\ \ \ \ -(N_e^S\!-\!1)\epsilon_e\!-\epsilon_fN_f^S\nonumber\\
&&=\epsilon_e(1\!-\!\frac{N_e^S}{N_e})+\epsilon_f(\frac{N_fN_e^S\!-\!N_f^SN_e}{NN_e})\nonumber\\
&&\geq{(1\!-\!\frac{N_e^S}{N_e})\frac{\epsilon_f}{N}\!\cdot\!\frac{N_f^SN_e\!-\!N_fN_e^S}{N_e\!-\!N_e^S}\!+\!\epsilon_f(\frac{N_fN_e^S\!-\!N_f^SN_e}{NN_e})}\nonumber\\
&&=\epsilon_f\!\left(\!\frac{N_f^SN_e\!-\!N_fN_e^S}{NN_e}\!+\!\frac{N_fN_e^S\!-\!N_f^SN_e}{NN_e}\!\right)\!\!=\!0.\nonumber
\end{eqnarray}
If $1\!\leq\!{N_e^S}\!=\!N_e$, due to $N_f^S\!\leq\!{N}_f\!<\!N$, we have
\begin{eqnarray}
\sum_{i\in\mathcal{S}}\phi_i\!-v(\mathcal{S})\!=\!\epsilon_f(\frac{N_f\!-\!N_f^S}{N})\!\geq\!{0}.\nonumber
\end{eqnarray}
Similarly, if $N_e^S\!=\!0$, it follows that
\begin{eqnarray}
\sum_{i\in\mathcal{S}}\phi_i\!-v(\mathcal{S})&&\!\!=\!(1\!-\!\frac{1}{N})\epsilon_fN_f^S\!-\epsilon_f(N_f^S\!-\!1)\nonumber\\
&&\!\!=\!\epsilon_f(1\!-\!\frac{N_f^S}{N})\!\geq\!{0}.\nonumber
\end{eqnarray}
Given the above, we show that $\sum_{i\in\mathcal{S}}\!\phi_i\!\geq\!{v(\mathcal{S})}$ holds for any $\mathcal{S}\!\subseteq\!{\mathcal{N}}$. Therefore, $\phi\!\in\!\mathrm{Core}(G)$. This completes the proof of Lemma~\ref{Lemma2}. \hfill $\square$ 

Lemma~\ref{Lemma2} presents the necessary and sufficient condition ensuring core stability of the Shapley value-based payoff allocation. Nevertheless, it involves checking all possible coalitions $\mathcal{S}\!\subseteq\!{\mathcal{N}}$, which can be computationally intensive. To address this complexity, Theorem~\ref{Theorem2} provides a more tractable sufficient condition.

\begin{theorem}\label{Theorem2}
Let $\phi\!=\![\phi_1,\dots,\phi_N]^{\top}\!\!\in\!\mathbb{R}^N$ be the Shapley value-based payoff allocation given in Proposition~\ref{Proposition1} with $N_f\!>\!0$ and $N_e\!>\!1$. If the following condition holds:
\begin{eqnarray}
\frac{\epsilon_e}{\epsilon_f}\geq\frac{N_f}{N},\label{Equ.28}
\end{eqnarray}
then the allocation $\phi$ lies in the core of the game $G$.
\end{theorem}
\noindent \textbf{Proof.}
By Lemma~\ref{Lemma2}, $\phi\!\in\!\mathrm{Core}(G)$ if and only if \eqref{Equ.25} holds for any coalition $\mathcal{S}\!\subseteq\!\mathcal{N}$ with $1\!\leq\!{N_e^S}\!<\!N_e$. To guarantee this inequality, we consider the worst case that maximizes the right-hand side. Since $N_f^S\!\leq\!{N_f}$, setting $N_f^S\!=\!N_f$ in \eqref{Equ.25} yields
\begin{eqnarray}
\frac{\epsilon_e}{\epsilon_f}\geq\frac{1}{N}\cdot\frac{N_f(N_e\!-\!N_e^S)}{N_e\!-\!N_e^S}\!=\!\frac{N_f}{N}.\nonumber
\end{eqnarray}
This concludes the proof of Theorem~\ref{Theorem2}. \hfill $\square$ 

Theorem~\ref{Theorem2} establishes sufficient conditions under which the Shapley value-based allocation is both fair and core-stable. It offers a more explicit and computationally efficient criterion for verifying core stability, relying only on the platoon composition ($N_f$, $N$) and the parameters ($\epsilon_e$, $\epsilon_f$). 

\begin{figure}[t]
\centering
\includegraphics[scale=0.35]{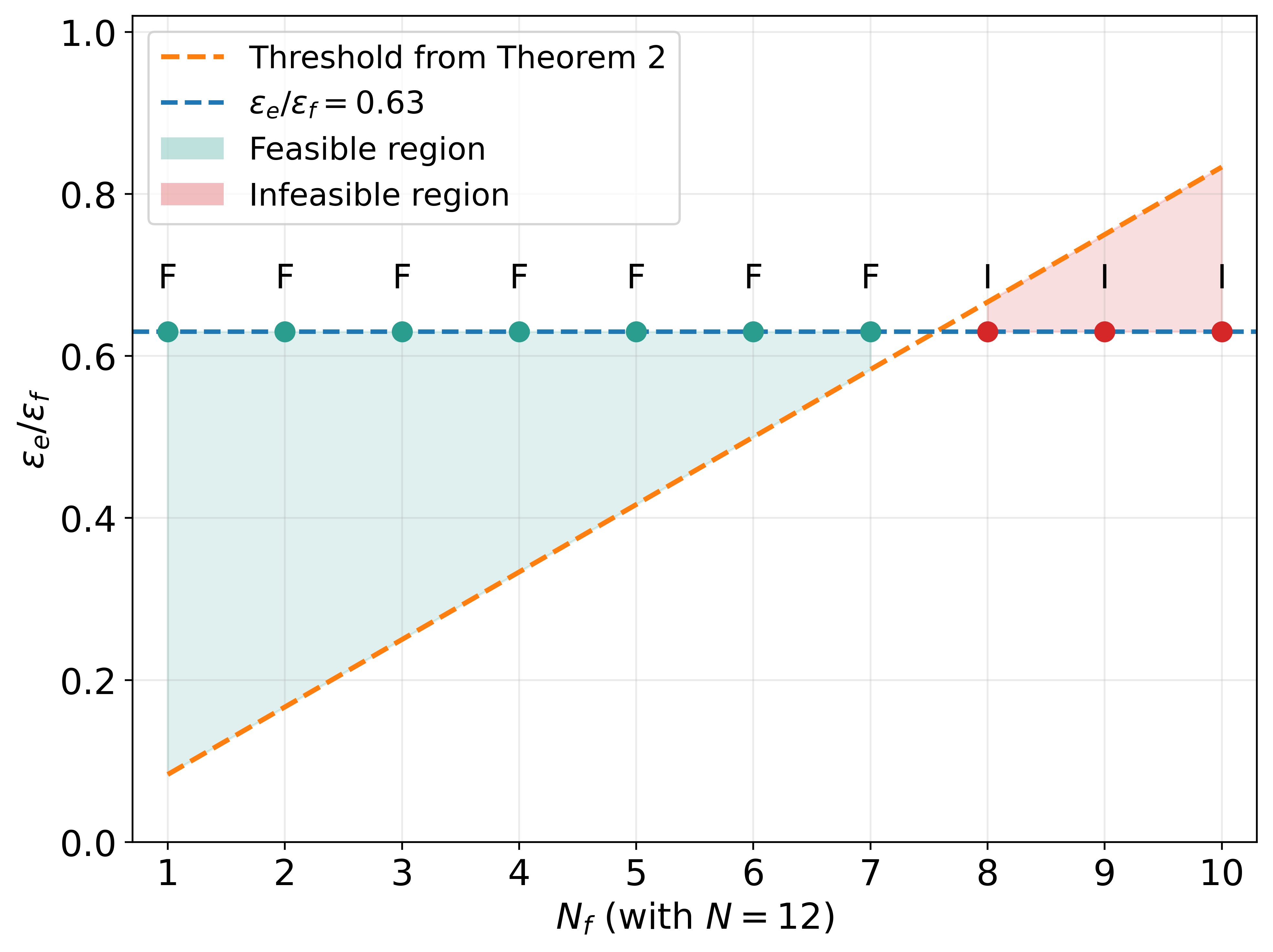}
\vspace{-5pt}
\caption{Core-stability regions of the Shapley value-based payoff allocation.}
\label{Fig.2}
\end{figure}

To improve the accessibility of the analytical results, Fig.~\ref{Fig.2} illustrates the parameter regimes under which the Shapley value-based payoff allocation is core-stable. Specifically, for a mixed-energy truck platoon with $N\!=\!12$ and $\epsilon_e/\epsilon_f\!=\!0.63$, the orange dotted line represents the boundary characterized in Theorem~\ref{Theorem2}, while the green and red regions denote the feasible and infeasible regions for core stability, respectively. Since $N_f$ takes integer values, the green and red markers indicate the corresponding feasible and infeasible parameter points.

\section{Trade-off between stability and fairness}\label{Section 5}
In this section, we deal with cases where the conditions in Theorem~\ref{Theorem2} are not fulfilled, and as a result, the Shapley value may fall outside the core. To address this limitation, we introduce an alternative allocation mechanism built upon the stable payoff structure presented in Theorem~\ref{Theorem1}, which guarantees core stability of the allocation while preserving the desirable fairness properties of the Shapley value to the greatest extent. 
\vspace{-5pt}

\begin{theorem}\label{Theorem3}
Consider the platoon coalitional game $G\!=\!(\mathcal{N}, v)$, where $v$ is defined in \eqref{Equ.2}, $N_f\!>\!0$ and $N_e\!>\!1$. Let $\phi\!=\![\phi_1,\dots,\phi_N]^{\top}\!\!\in\!\mathbb{R}^N$ be the Shapley value-based payoff allocation given by \eqref{Equ.18} and $x\!=\![x_1,\dots,x_N]^{\top}\!\!\in\!\mathbb{R}^N$ be any feasible allocation. Define the mean relative deviation of $x$ from $\phi$ by
\vspace{-5pt}
\begin{eqnarray}
\Delta_{\phi}(x)\!:=\!\frac{1}{N}\sum_{i\in\mathcal{N}}\left|\frac{\phi_i\!-\!x_i}{\phi_i}\right|.\label{Equ.29}
\end{eqnarray}
If condition \eqref{Equ.28} is not satisfied, define $x$ by \eqref{Equ.7} with 
\vspace{-3pt}
\begin{eqnarray}
\xi^*\!=\!\frac{\epsilon_e}{\epsilon_e(N_e\!-\!1)\!+\!\epsilon_fN_f}.\label{Equ.30}
\end{eqnarray}
Then: 
(i) The resulting payoff allocation $x(\xi^{*})$ lies in the core of game $G$ with $\Delta_{\phi}(x(\xi))$ minimized, i.e.,
\vspace{-3pt}
\begin{eqnarray}
&&\xi^{*}=\arg\min\limits_{0<\xi<{1}}\Delta_{\phi}(x(\xi))\nonumber\\
&& \mathrm{s.\,t.}\quad x(\xi)\!\in\!\mathrm{Core}(G).\nonumber
\end{eqnarray}
(ii) $\Delta_{\phi}(x(\xi^{*}))\!<\!1$.
\end{theorem}
\vspace{-1pt}

\begin{figure*}[t]
    \centering
    \subfigure[Mean relative deviation in different $\epsilon_e/\epsilon_f$.]{
        \includegraphics[width=0.425\linewidth]{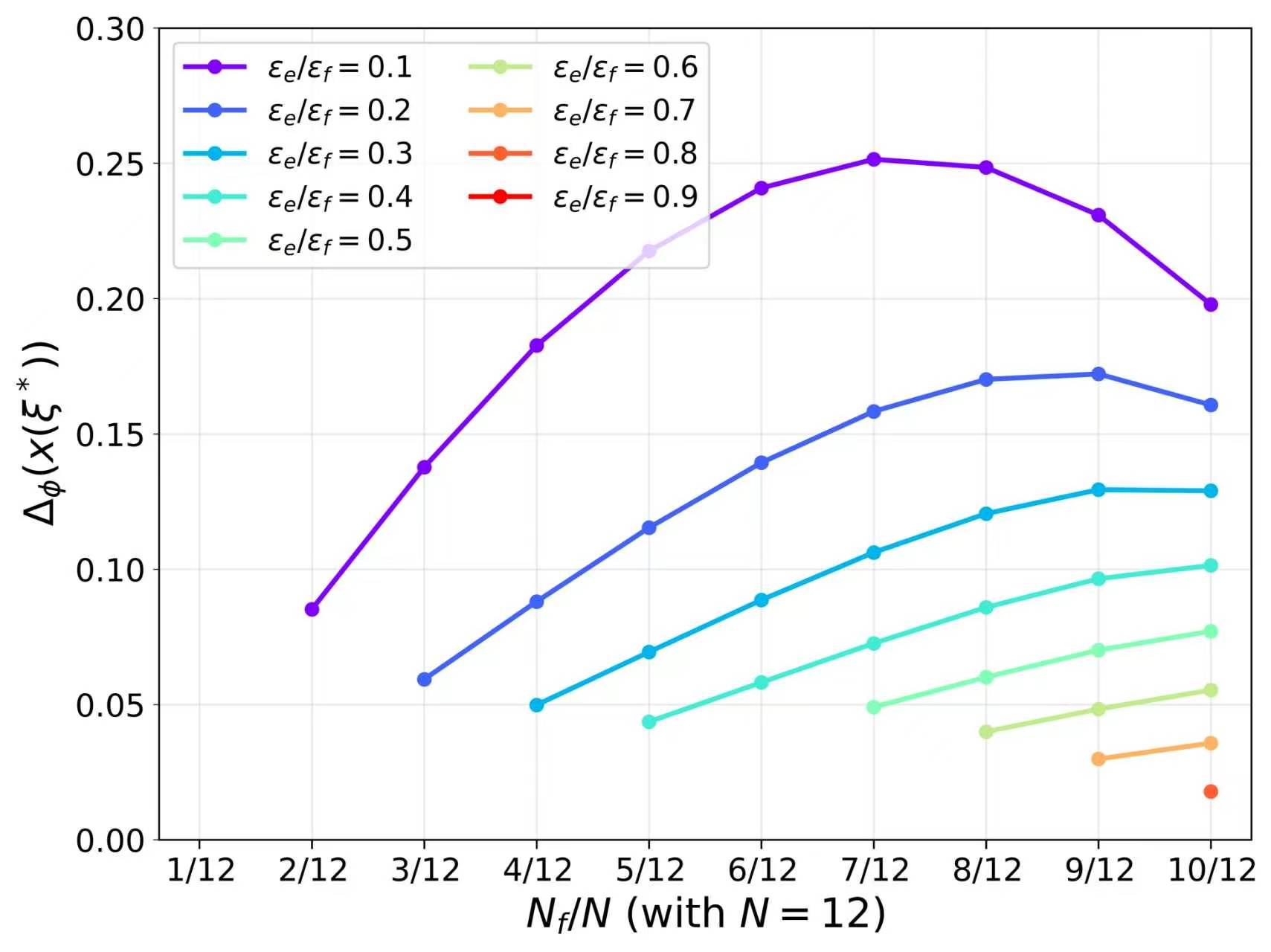}
        \label{Fig.3a}
    }
    \subfigure[Distribution of $\Delta_{\phi}(x(\xi^*))$.]{
        \includegraphics[width=0.455\linewidth]{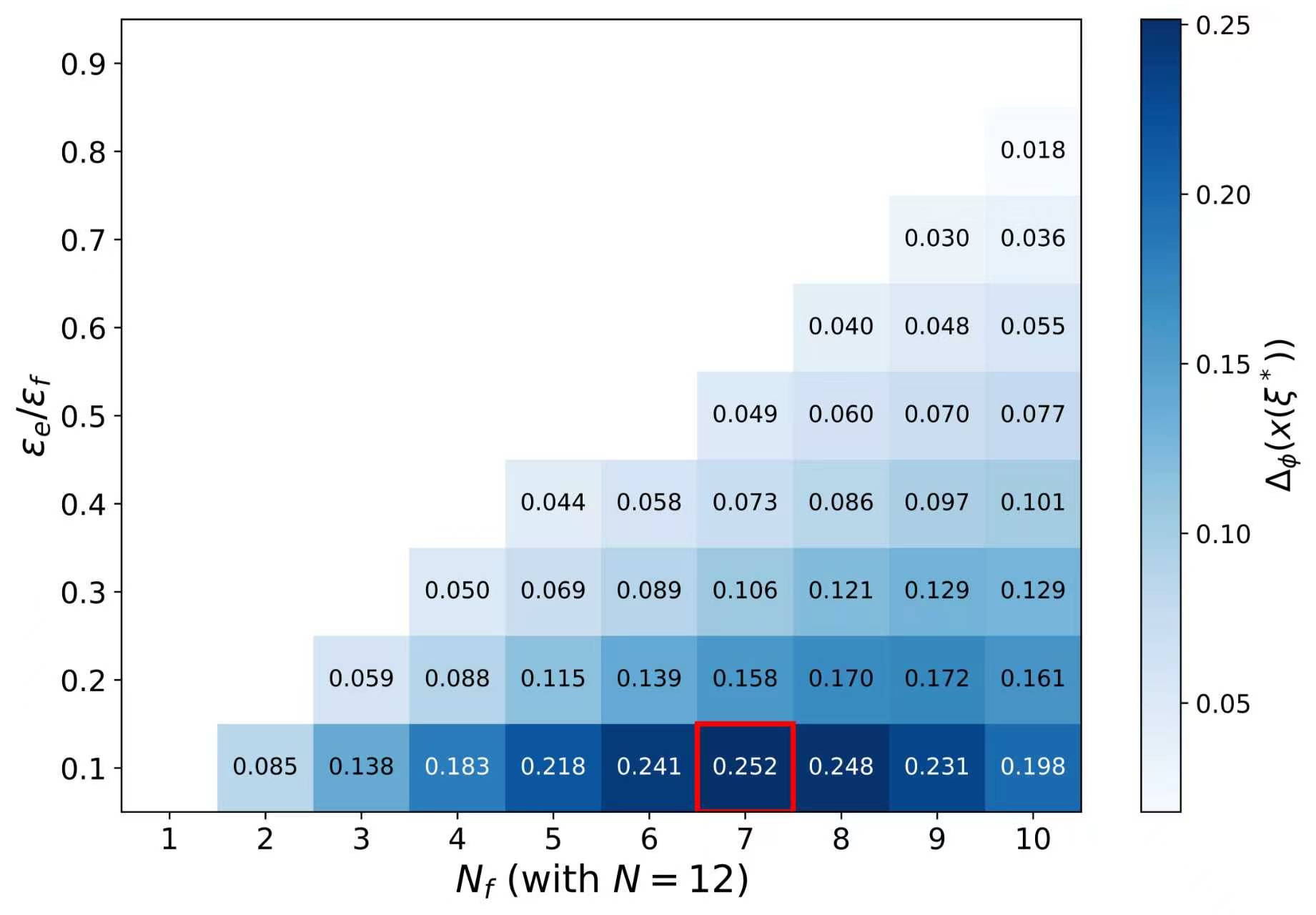}
        \label{Fig.3b}
    }
    \caption{Mean relative deviation $\Delta_{\phi}(x(\xi^{*}))$ under representative parameter settings.}
    \label{Fig.3}
\end{figure*}

\noindent \textbf{Proof.} (i) From Theorem~\ref{Theorem1}, the payoff allocation defined in~\eqref{Equ.7} with $\xi$ given by \eqref{Equ.30} lies in the core of game $G$. Recall that $\xi$ denotes the design parameter in the core-stable payoff family introduced in Theorem~\ref{Theorem1}. Below, we will show that $\xi^*\!=\!\frac{\epsilon_e}{\epsilon_e(N_e-1)+\epsilon_fN_f}$ is the optimal value of $\xi$ within its admissible range that minimizes $\Delta_{\phi}(x(\xi))$.
\vspace{-3pt}

Denote as $x(\xi)$ the payoff allocation defined by \eqref{Equ.7}, where $0\!<\!\xi\!\leq\!{\frac{\epsilon_e}{\epsilon_e(N_e-1)+\epsilon_fN_f}}$. As $N_e\!>\!0$, let $\phi_L\!\in\!\phi$ and $x_L(\xi)\!\in\!{x}$ be the payoff of the leader in the two allocation schemes with $L\!\in\!\mathcal{N}_e$. For simplicity, we denote by $\phi_e$ and $\phi_f$ the payoff of truck $i\!\in\!\mathcal{N}_e$ and truck $i\!\in\!\mathcal{N}_f$, respectively. By \eqref{Equ.29} and \eqref{Equ.7}, it follows that
\begin{eqnarray}
&&\Delta_{\phi}(x(\xi))\nonumber\\
=\!&&\frac{1}{N}\!\sum_{i\in\mathcal{N}}\left|\frac{\phi_i\!-\!x_i(\xi)}{\phi_i}\right|\nonumber\\
=\!&&\frac{1}{N}\!\bigg(\frac{|\phi_L\!-\!x_L(\xi)|}{\phi_L}\!+\!\!\!\!\sum_{i\in{\mathcal{N}}_e\setminus\{L\}}\!\!\!\!\!\!\!\frac{|\phi_i\!-\!x_i(\xi)|}{\phi_i}\!+\!\!\sum_{i\in{\mathcal{N}}_f}\!\!\frac{|\phi_i\!-\!x_i(\xi)|}{\phi_i}\bigg)\nonumber\\
=\!&&\frac{1}{N}\!\bigg(\frac{|\phi_L\!-\!x_L(\xi)|}{\phi_L}+(N_e\!-\!1)\frac{|\phi_e\!-\!(1\!-\!\xi)\epsilon_e|}{\phi_e}\nonumber\\
&&\ \ \ \ \ \ \ +N_f\frac{|\phi_f\!-\!(1\!-\!\xi)\epsilon_f|}{\phi_f}\bigg),\label{Equ.31}
\end{eqnarray}
where, by \eqref{Equ.2} and \eqref{Equ.18}, we have
\begin{eqnarray}
x_L(\xi)&&\!=\xi v(\mathcal{N})\!=\xi\big((N_e\!-\!1)\epsilon_e\!+\!N_f\epsilon_f\big),\nonumber\\
\phi_L&&\!=\phi_e\!=(1\!-\!\frac{1}{N_e})\epsilon_e\!+\!\frac{N_f}{NN_e}\epsilon_f,\nonumber\\
\phi_f&&\!=(1\!-\!\frac{1}{N})\epsilon_f.\nonumber
\end{eqnarray}
Violation of condition \eqref{Equ.28} implies that $\epsilon_fN_f\!>\!\epsilon_eN$. In addition, as $0\!<\!\xi\!\leq\!\frac{\epsilon_e}{\epsilon_e(N_e-1)+\epsilon_fN_f}$, we obtain that
\begin{eqnarray}
\phi_L\!-\!x_L(\xi)&&\!\!=\!(1\!-\!\frac{1}{N_e})\epsilon_e\!+\!\frac{N_f}{NN_e}\epsilon_f\!-\!\xi\big((N_e\!-\!1)\epsilon_e\!+\!N_f\epsilon_f\big)\nonumber\\
&&\!\!\geq\!{(1\!-\!\frac{1}{N_e})\epsilon_e\!+\!\frac{N_f}{NN_e}\epsilon_f-\epsilon_e}\nonumber\\
&&\!\!=\!\frac{1}{N_e}(\frac{\epsilon_fN_f}{N}\!-\epsilon_e)\nonumber\\
&&\!\!>\!\frac{1}{N_e}(\frac{\epsilon_eN}{N}\!-\epsilon_e)\!=\!0.\label{Equ.32}
\end{eqnarray}
Due to $\frac{\epsilon_fN_f}{\epsilon_eN}\!>\!1$, the deviation of $\phi_e$ from $(1\!-\!\xi)\epsilon_e$ is expressed as
\begin{eqnarray}
\phi_e\!-\!(1\!-\!\xi)\epsilon_e&&\!=(1\!-\!\frac{1}{N_e})\epsilon_e\!+\!\frac{N_f}{NN_e}\epsilon_f\!-\!(1\!-\!\xi)\epsilon_e\nonumber\\
&&\!=(\xi\!-\!\frac{1}{N_e})\epsilon_e\!+\!\frac{N_f}{NN_e}\epsilon_f\nonumber\\
&&\!=\frac{\epsilon_e}{N_e}(\xi{N_e}\!-\!1\!+\!\frac{N_f\epsilon_f}{N\epsilon_e})\!>\!0.\label{Equ.33}
\end{eqnarray}
Since $\xi\!\leq\!{\frac{\epsilon_e}{\epsilon_e(N_e-1)+\epsilon_fN_f}}$ and $N_e\!\geq\!1$, the following inequality holds:
\begin{eqnarray}
\xi\leq{\frac{1}{(N_e\!-\!1)\!+\!N\!\cdot\!{\frac{\epsilon_fN_f}{N\epsilon_e}}}}<\frac{1}{(N_e\!-\!1)\!+\!N}\leq\frac{1}{N}.\label{Equ.34}
\end{eqnarray}
By \eqref{Equ.34}, one has
\begin{eqnarray}
\phi_f\!-\!(1\!-\!\xi)\epsilon_f\!=\!(1\!-\!\frac{1}{N})\epsilon_f\!-\!(1\!-\!\xi)\epsilon_f\!=\!(\xi\!-\!\frac{1}{N})\epsilon_f\!<\!0.\label{Equ.35}
\end{eqnarray}
Substituting \eqref{Equ.32}, \eqref{Equ.33}, \eqref{Equ.35} and $x_L(\xi)$ into \eqref{Equ.31} yields
\begin{eqnarray}
\Delta_{\phi}(x(\xi))&&\!\!=\!\frac{1}{N}\!\bigg(\!1\!-\!\frac{x_L(\xi)}{\phi_L}+(N_e\!-\!1)\big(1\!-\!\frac{(1\!-\!\xi)\epsilon_e}{\phi_e}\big)\nonumber\\
&&\ \ \ \ \ \ \ \ \  +N_f\big(\frac{(1\!-\!\xi)\epsilon_f}{\phi_f}\!-\!1\big)\!\bigg)\nonumber\\
&&\!\!\!=\!\frac{1}{N}\!\bigg(\!N_e\!-\!N_f-\frac{\xi\big((N_e\!-\!1)\epsilon_e\!+\!N_f\epsilon_f\big)}{\phi_e}\nonumber\\
&&\ \ \ \ \ \ \ \  -(N_e\!-\!1)\frac{(1\!-\!\xi)\epsilon_e}{\phi_e}+N_f\frac{(1\!-\!\xi)\epsilon_f}{\phi_f}\!\bigg)\nonumber\\
&&\!\!\!=\!\frac{1}{N}\!\bigg(\!\!N_e\!-\!N_f\!-\!\frac{N_f\epsilon_f\xi\!+\!(N_e\!\!-\!\!1)\epsilon_e}{\phi_e}\!+\!\frac{(1\!\!-\!\xi)\epsilon_fN_f}{\phi_f}\!\!\bigg)\nonumber\\
&&\!\!\!=\!\frac{1}{N}\!\bigg(\!N_e\!-\!N_f-\frac{(N_e\!\!-\!1)\epsilon_e}{\phi_e}+\frac{\epsilon_fN_f}{\phi_f}\nonumber\\
&&\ \ \ \ \ \ \ \  -N_f\epsilon_f({\frac{1}{\phi_e}\!+\!\frac{1}{\phi_f}})\xi\!\bigg).\label{Equ.36}
\end{eqnarray}
This allows us to derive that
\begin{eqnarray}
\frac{d}{d\xi}\Delta_{\phi}(x(\xi))\!=\!-\frac{N_f\epsilon_f}{N}(\frac{1}{\phi_e}\!+\!\frac{1}{\phi_f})\!<\!0,\nonumber
\end{eqnarray}
where $\phi_i\!>\!0$, $\forall{i\!\in\!\mathcal{N}}$. Consequently, $\Delta_{\phi}(x(\xi))$ decreases with the increase of $\xi$ and the minimum value is achieved at $\xi^*\!=\!\frac{\epsilon_e}{\epsilon_e(N_e-1)+\epsilon_fN_f}$. 

(ii) By \eqref{Equ.36}, since $\xi\!>\!0$, $N_e\!-\!1\!\geq\!0$, $N\!\geq\!{2}$, and $\phi_f\!=\!(1\!-\!\frac{1}{N})\epsilon_f$, one has
\begin{eqnarray}
\Delta_{\phi}(x(\xi^{*}))&&\!<\!\frac{1}{N}(N_e\!-\!N_f\!+\!\frac{\epsilon_fN_f}{\phi_f})\!=\!\frac{1}{N}(N_e\!-\!N_f\!+\!\frac{NN_f}{N\!-\!1})\nonumber\\
&&=\!\frac{1}{N}(N_e\!+\!\frac{N_f}{N\!-\!1})\!\leq\!\frac{1}{N}(N_e\!+\!N_f)\!=\!1.\nonumber
\end{eqnarray}
This completes the proof of Theorem~\ref{Theorem3}. \hfill $\square$

In Theorem~\ref{Theorem3}, the Shapley value-based payoff allocation $\phi$ serves as a fair allocation benchmark, and $x(\xi^*)$, defined in Theorem~\ref{Theorem1} with $\xi^*\!=\!\frac{\epsilon_e}{\epsilon_e(N_e-1)+\epsilon_fN_f}$, provides a core-stable allocation that minimizes the mean relative deviation $\Delta_{\phi}(x(\xi))$ over $\xi$. The inequality $\Delta_{\phi}(x(\xi^*))\!<\!1$ indicates that, on average, each truck’s payoff deviates by less than $100\%$ from its fair share under the Shapley value, ensuring a bounded level of unfairness. The results in Theorem~\ref{Theorem3} guarantee that no truck is substantially over- or under-compensated, thereby enabling a stable and reasonably fair distribution of benefits even when perfect adherence to the Shapley value is infeasible. 

To provide practical insight into how often and how severely the Shapley value-based payoff allocation falls outside the core, Fig.~\ref{Fig.3} shows the mean relative deviation $\Delta_{\phi}(x(\xi^{*}))$ under representative parameter settings, with $N\!=\!12$, $N_f\!=\!1,\dots,10$, and $\epsilon_e/\epsilon_f\!=\!0.1,0.2,\dots,0.9$. As shown in Fig.~\ref{Fig.3a}, the deviation generally increases as $\epsilon_e/\epsilon_f$ decreases and is most pronounced in moderately FPT-dominated fleets. Fig.~\ref{Fig.3b} illustrates the distribution of $\Delta_{\phi}(x(\xi^{*}))$ over the tested parameter grid, with the worst-case point highlighted in red. These results indicate that the Shapley value-based payoff allocation is more likely to lose core-stability when $\epsilon_e/\epsilon_f$ is small and the fleet is FPT-dominated, while the deviation from the Shapley value remains moderate under the proposed core-stable allocation $x(\xi^{*})$.

\begin{remark}\label{Remark6}
   Based on the above results, we are now able to answer the second question raised in Section~\ref{Section2.2}. Specifically, under the optimal coalition structure $\mathcal{N}$, if condition \eqref{Equ.28} holds, then employing the Shapley value-based allocation given in \eqref{Equ.18} ensures both core stability and fairness. However, for scenarios where condition \eqref{Equ.28} is not satisfied, one can construct the payoff allocation $x(\xi^*)$ as defined in \eqref{Equ.7} with $\xi^*$ given by~\eqref{Equ.30} to achieve minimal unfairness of the allocation while preserving core stability.
\end{remark}

\begin{remark}
This paper adopts a type-based benefit model where the follower benefit parameters $\epsilon_f$ and $\epsilon_e$ are assumed to be known and uniform within each truck type. This approximation enables the characteristic function, as well as the associated fairness and stability properties, to be characterized explicitly in closed form. In practice, however, follower benefits may vary across trucks due to differences in truck characteristics, operating conditions, or positions within a platoon, while the corresponding parameters may also be uncertain or privately held. Under small within-type heterogeneity, allocations with a sufficient stability margin are expected to remain core-stable, whereas those near the core boundary could be more sensitive to parameter variations. Accounting for such heterogeneity and uncertainty would require a more general estimated or robust formulation of the characteristic function. Extending the proposed framework in these directions is an important topic for future work.
\end{remark}

\section{Numerical examples}\label{Section 6}
This section presents numerical studies to validate the main theoretical results. We start with assessing the superadditivity of the platoon coalitional game $G$ and the core stability of the allocation scheme proposed in Theorem~\ref{Theorem1}. Next, we evaluate the Shapley value-based allocation presented in Proposition~\ref{Proposition1}, as well as its core stability under the conditions in Theorem~\ref{Theorem2}. Finally, the deviation-minimized allocation given in Theorem~\ref{Theorem3} is validated. The code implementation is available online.\footnote{See code at: {https://github.com/kth-tingbai/Stable-and-Fair-Platooning-Benefit-Allocation}}

\textit{Parameter settings:} We consider a travel distance of $300$~km between logistics hubs $O$ and $D$. Consistent with the theoretical formulation, where the trip distance is normalized to $1$, with $\epsilon_f$ and $\epsilon_e$ denoting the monetary savings per unit travel distance, the total platooning benefit over the actual trip is obtained by linearly scaling the unit-distance savings with the travel distance. The benefit of platooning is quantified as the monetary savings resulting from reduced energy consumption, assuming each following truck achieves a $10\%$ reduction. Truck-related parameters are set based on data from Scania~\citep{ElectricTruck}. Specifically, all trucks are assumed to travel at a constant speed of $80$ km/h. The fuel consumption for FPTs is set as $0.4$~L/km, while ETs consume $1.34$~kWh/km. Based on current energy prices, the monetary savings per following truck from platooning are approximately $\epsilon_f\!=\!0.07$~\texteuro/km for FPTs and $\epsilon_e\!=\!0.048$~\texteuro/km for ETs. Due to safety and technical constraints, the maximum platoon size is set to $N\!\leq\!M\!=\!15$ based on the existing truck-platooning studies~\citep{yang2025optimization}.

\begin{table}[t]
    \centering
    \caption{Platoon formation configurations (coalitions) and the resulting total platooning benefit.}
    \label{tab:runtime_comparison}
    \scriptsize  
    \renewcommand{\arraystretch}{1.15}
    \setlength{\tabcolsep}{15pt}  

    \newcommand{\thintoprule}{\specialrule{0.8pt}{0pt}{1pt}}
    \begin{tabular}{c p{2.75cm} c}
        \thintoprule
        \!\!\textbf{Case}\!\!\! & \!\!\!\textbf{Coalitions}\!\!\!\!\!\! & \!\!\!\!\!\textbf{Total Benefit} [\texteuro]\!\!\! \\
        \midrule
       \rowcolor{blue!10} 1 & $(EEDDD)$ & $77.4$\\
        2 & $(EE),(DDD)$ & $56.4$\\
        3 & $(ED),(EDD)$ & $63.0$\\
        4 & $(DD),(EED)$ & $56.4$\\
        5 & $(E),(EDDD)$ & $63.0$\\
        6 & $(D),(EEDD)$ & $56.4$\\
        7 & $(E),(E),(DDD)$ & $42.0$ \\
        8 & $(E),(D),(EDD)$ & $42.0$ \\
        9 & $(D),(D),(EED)$ & $35.4$ \\
        10 & $(E),(DD),(ED)$ & $42.0$ \\
        11 & $(D),(ED),(ED)$ & $42.0$ \\
        12 & $(D),(DD),(EE)$ & $35.4$ \\
        13 & $(DD),(E),(E),(D)$ & $21.0$ \\
        14 & $(ED),(E),(D),(D)$ & $21.0$ \\
        15 & $(EE),(D),(D),(D)$ & $14.4$ \\
        16 & $(E)$,$(E)$,$(D)$,$(D)$,$(D)$ & $0.0$\\
        \bottomrule
    \end{tabular}\label{Table1}
\end{table}

\begin{figure*}[t]
  \centering
  \subfigure[Case 1: $\epsilon_f\!=\!0.07$, $\epsilon_e\!=\!0.048$.]{
    \includegraphics[width=0.45\linewidth]{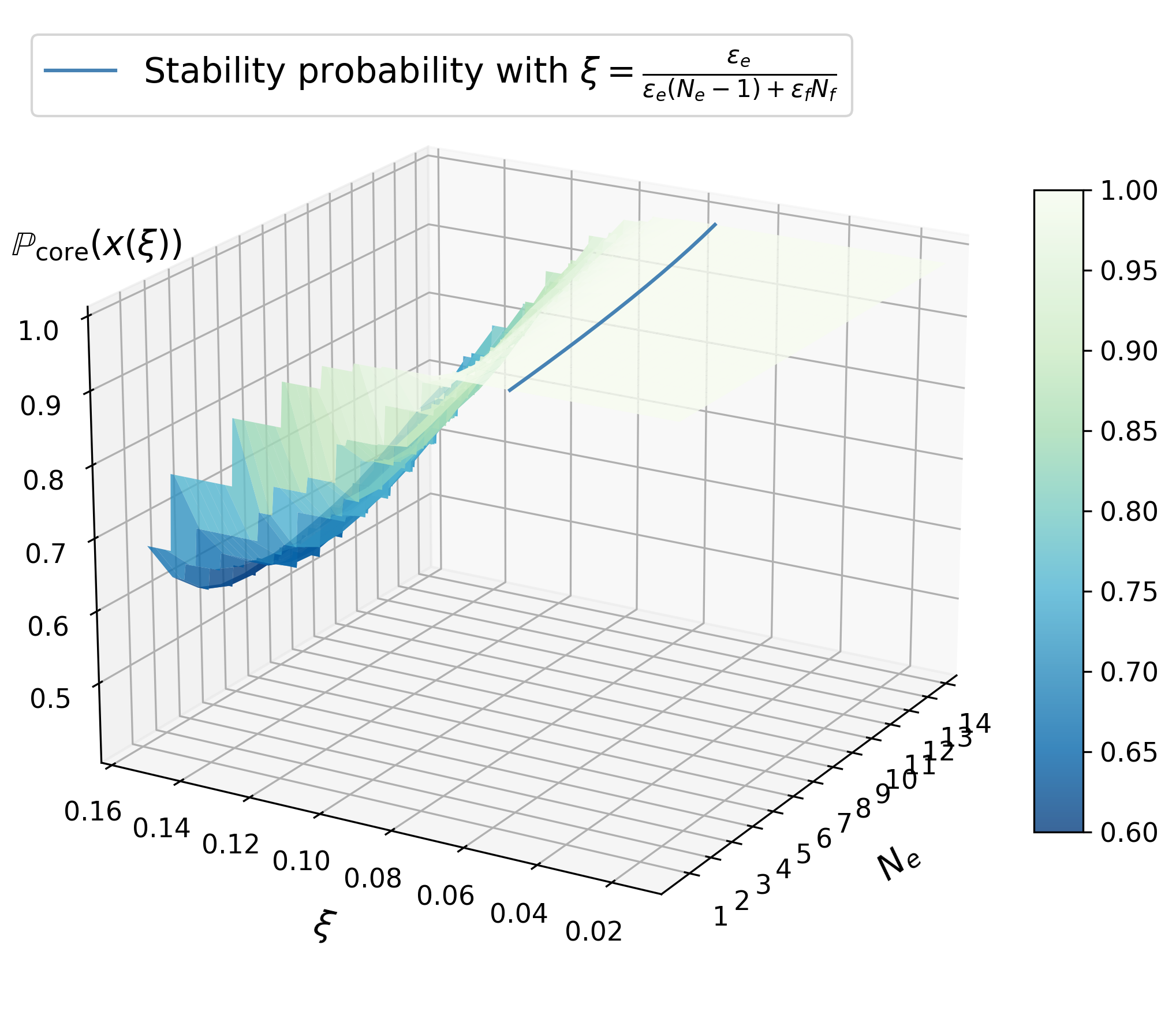}
    \label{Fig.4a}
 }
  \hfill
  \subfigure[Case 2: $\epsilon_f\!=\!0.10$, $\epsilon_e\!=\!0.048$.]{
    \includegraphics[width=0.45\linewidth]{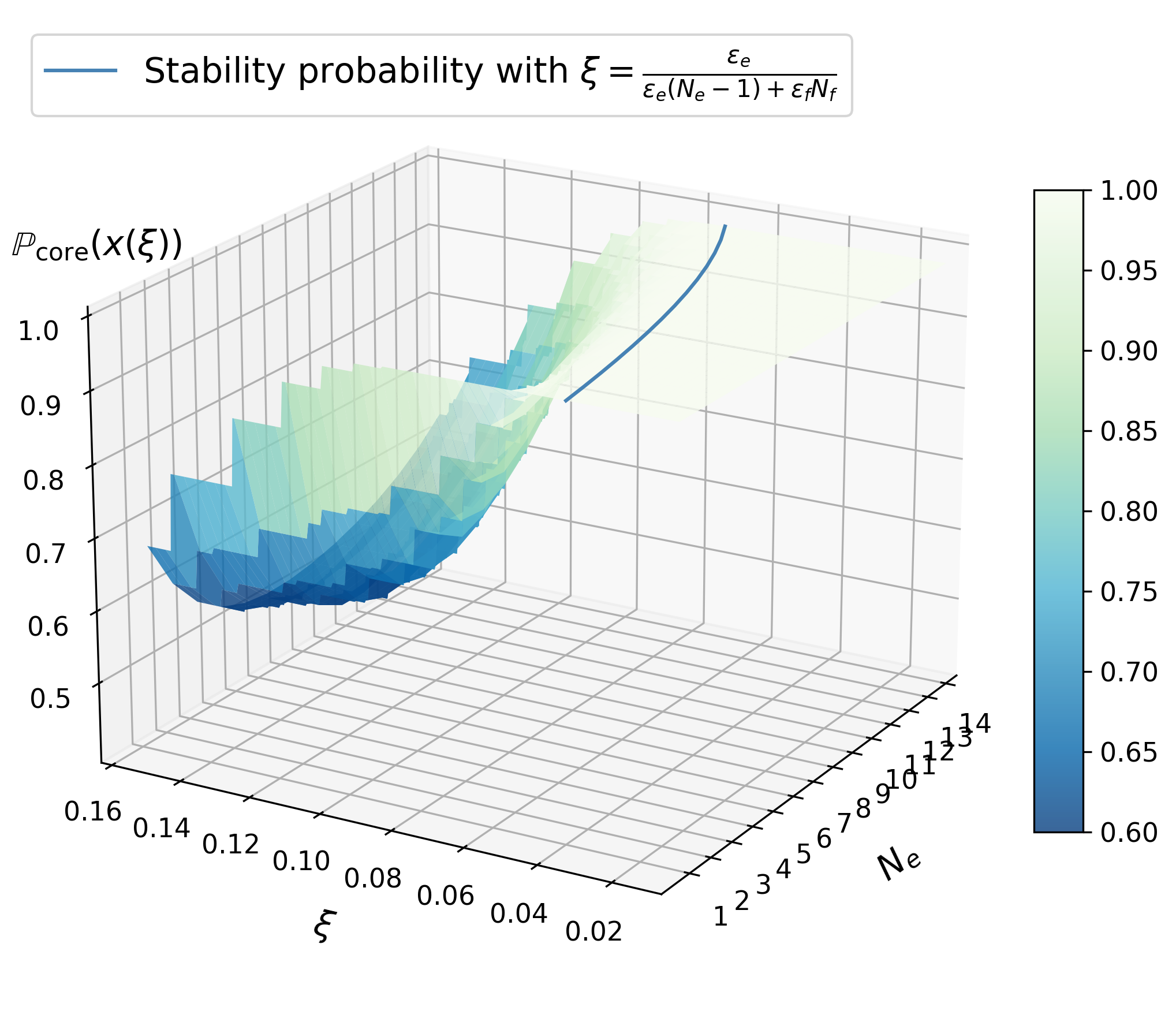}
    \label{Fig.4b}
  }
  \vskip\baselineskip
  \subfigure[Case 3: $\epsilon_f\!=\!0.13$, $\epsilon_e\!=\!0.048$.]{
    \includegraphics[width=0.45\linewidth]{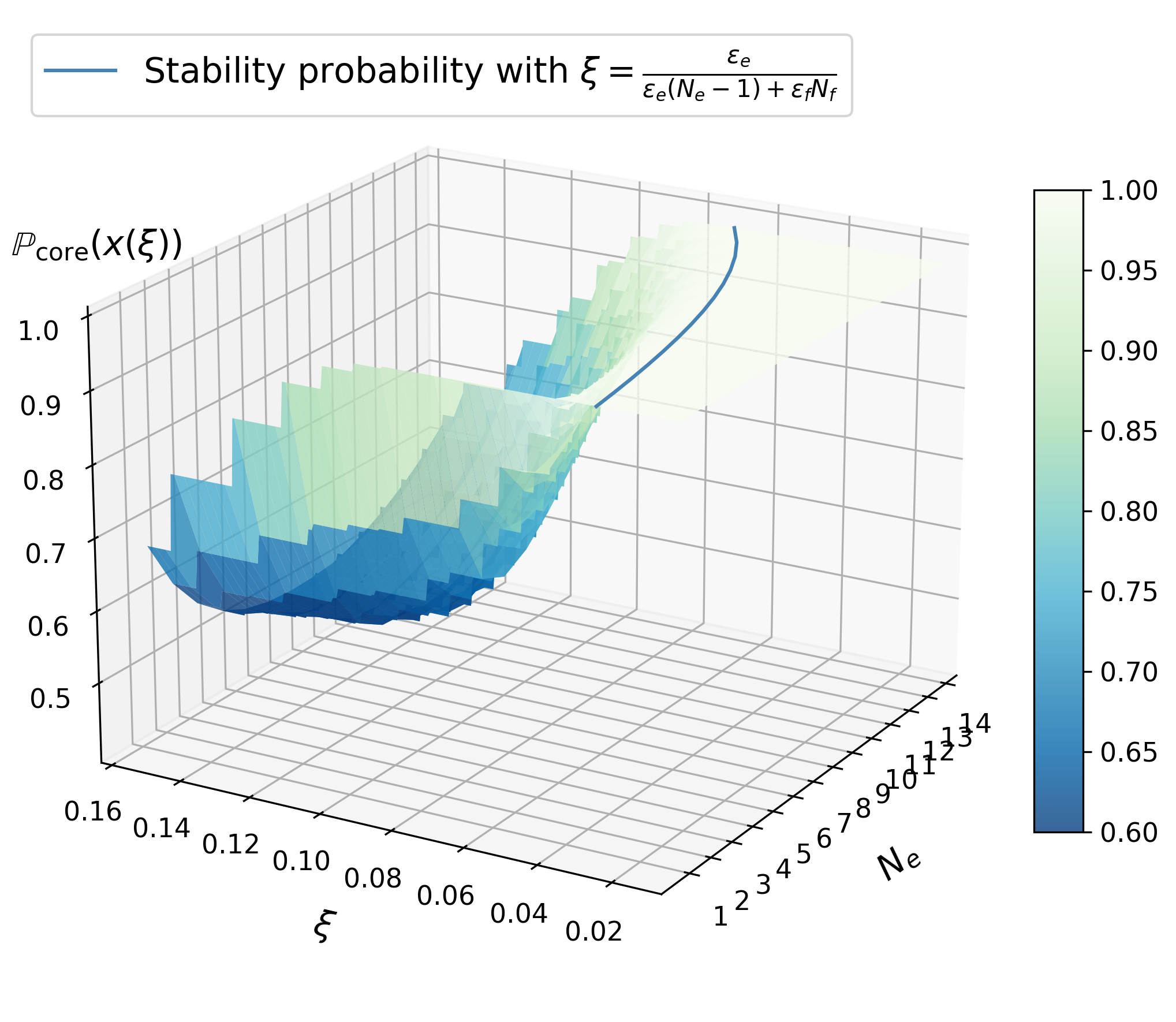}
    \label{Fig.4c}
  }
  \hfill
  \subfigure[Case 4: $\epsilon_f\!=\!0.07$, $\epsilon_e\!=\!0.005$.]{
    \includegraphics[width=0.45\linewidth]{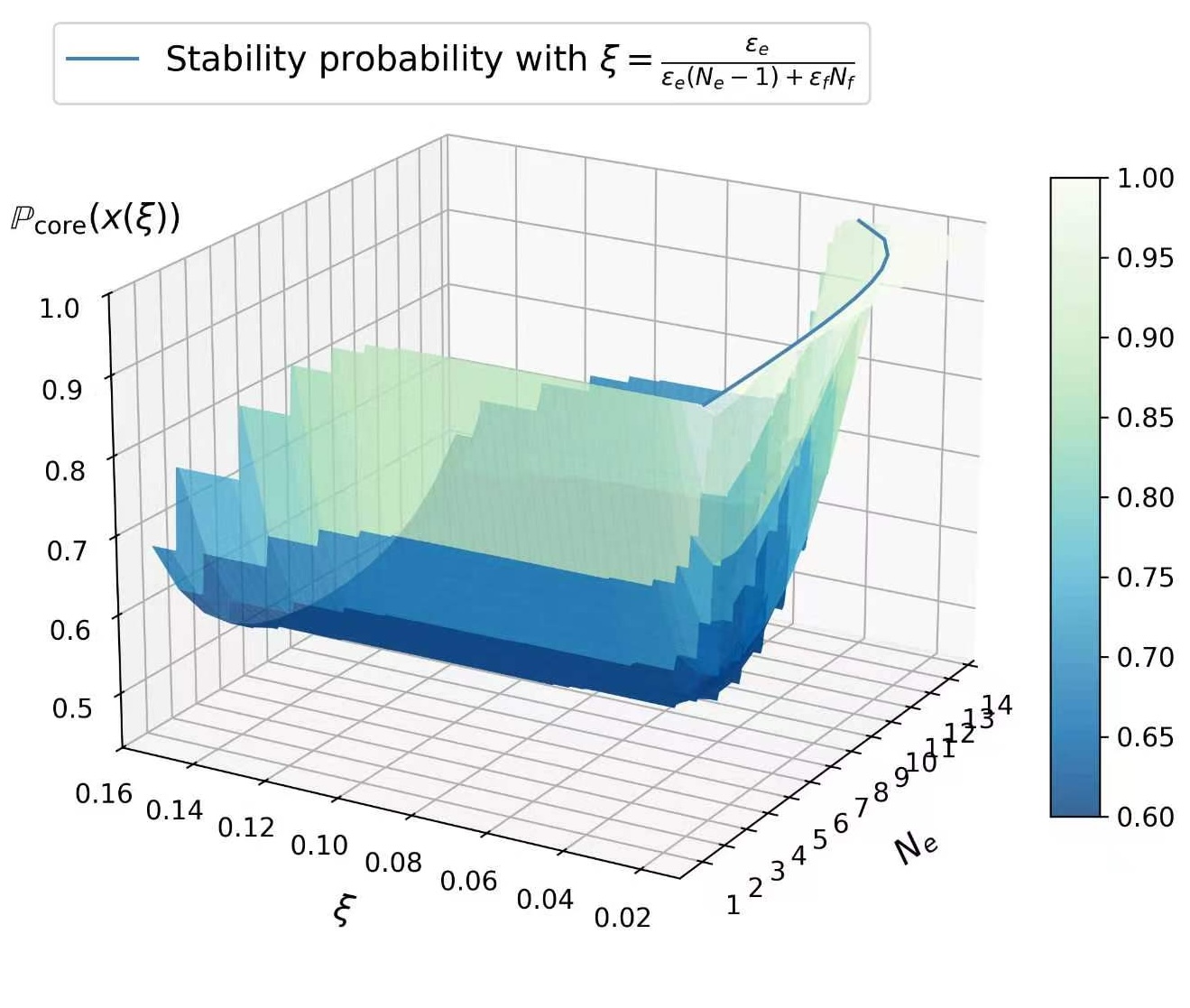}
    \label{Fig.4d}
  }
  \caption{Stability probability of allocation $x(\xi)$ in heterogeneous platoons across different cases.}
  \label{Fig.4}
\end{figure*}

\subsection{Performance of the stable allocation}\label{Section6.1}
The optimality of the grand coalition is first verified. Table~\ref{Table1} presents the total platooning benefits for all possible coalitions among five trucks, comprising $2$ ETs (denoted as E) and $3$ FPTs (denoted as D), as illustrated in Fig.~\ref{Fig.1}. The results show that the grand coalition $(EEDDD)$ yields the highest total benefit, thus serving as the optimal coalition. This verifies the superadditivity property of game $G\!=\!(\mathcal{N},v)$, as revealed in Lemma~\ref{Lemma1}.

To evaluate the core stability of the payoff allocation scheme proposed in Theorem~\ref{Theorem1}, we define the \textit{stability probability} of an allocation $x(\xi)$ as
\begin{eqnarray}
\mathbb{P}_{\mathrm{core}}(x(\xi))&&\!\!:=\!\mathbb{P}\big(x(\xi)\!\in\!\mathrm{Core}(G)\big)\nonumber\\
&&\!=\!1\!-\!\frac{\left|\{\mathcal{S}\!\subset\!{\mathcal{N}}\!\!:\!v(\mathcal{S})\!>\!\!\sum_{i\in{\mathcal{S}}}\!x_i(\xi),\mathcal{S}\!\neq\!{\emptyset}\}\right|}{2^N\!\!-\!2}\!,\label{Equ.37}
\end{eqnarray}
where the numerator represents the number of coalitions that would gain more by deviating from the grand coalition, while the denominator is the total number of possible coalitions, excluding the empty set and the grand coalition itself. The characteristic function $v(\mathcal{S})$, as defined in \eqref{Equ.2}, denotes the maximum platooning benefit that coalition $\mathcal{S}$ can achieve under optimal leader selection. By definition, $\mathbb{P}_{\mathrm{core}}(x(\xi))$ represents the probability that all trucks remain in the grand coalition $\mathcal{N}$. A value of $\mathbb{P}_{\mathrm{core}}(x(\xi))\!=\!1$ indicates that the allocation $x(\xi)$ lies within the core of game $G$, meaning no subset of trucks has an incentive to deviate. In contrast, smaller values of $\mathbb{P}_{\mathrm{core}}(x(\xi))$ reflect a higher probability of deviation, indicating lower stability of the platoon formation. 

\begin{figure}[t]
\centering
\includegraphics[scale=0.416]{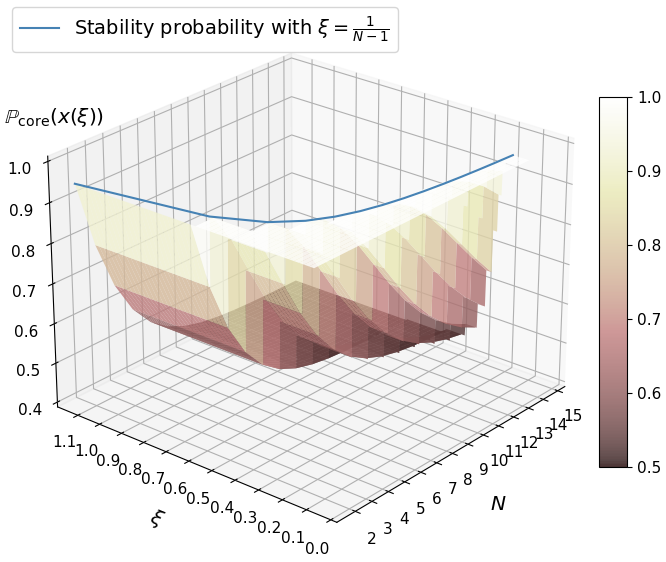}
\vspace{-5pt}
\caption{Stability probability of allocation $x(\xi)$ in homogeneous platoons, with $N\!=\!N_f$ and $\epsilon_f\!=\!0.07$.}
\label{Fig.5}
\end{figure}

Fig.~\ref{Fig.4a} shows the stability probability for different allocations $x(\xi)$ defined in \eqref{Equ.7}, with $\xi$ ranging from $0.005$ to $0.15$. The parameters are set as $\epsilon_f\!=\!0.07$~\texteuro/km, $\epsilon_e\!=\!0.048$~\texteuro/km, and $N\!=\!15$, while the number of ETs (i.e., $N_e$) varies from $1$ to $14$. It shows that, when $0\!<\!\xi\!\leq\!{\frac{\epsilon_e}{\epsilon_e(N_e-1)+\epsilon_fN_f}}$, the stability probability of $x(\xi)$ equals $1$, ensuring that $x(\xi)\!\in\!\mathrm{core}(G)$. This observation validates the condition established in Theorem~\ref{Theorem1}. We further incorporate carbon emission reductions from FPTs during platoon formation, which enhances the benefits of platooning, especially when carbon pricing mechanisms (e.g., carbon taxes) are applied. Fig.~\ref{Fig.4b} and Fig.~\ref{Fig.4c} evaluate the allocation stability probability under increased values of $\epsilon_f$, while Fig.~\ref{Fig.4d} verifies the stability probability under a stress-test setting with a very low value of $\epsilon_e$. The simulation across different scenarios validates the theoretical findings in Theorem~\ref{Theorem1}. Moreover, the results in Fig.~\ref{Fig.4} show that as the disparity between $\epsilon_f$ and $\epsilon_e$ increases, the feasible range of $\xi$ required to maintain the core stability of $x(\xi)$ becomes narrower. The validation for the special case in which all trucks are homogeneous is given in Fig.~\ref{Fig.5}.

\subsection{Evaluation of the Shapley value-based allocation}\label{Section6.2}
\begin{figure}[t]
\centering
\includegraphics[scale=0.225]{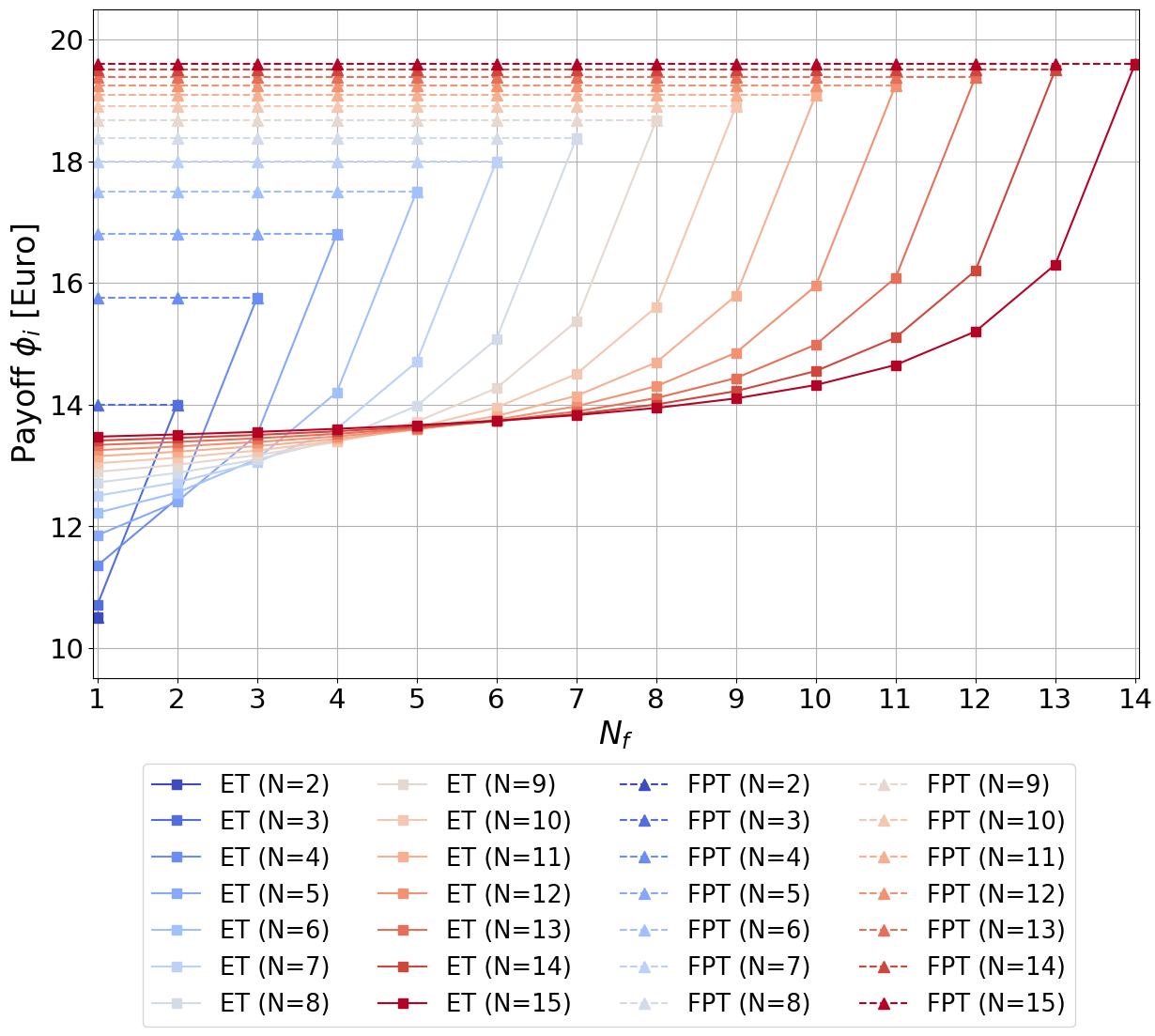}
\vspace{-2pt}
\caption{Shapley value-based payoff allocation.}
\label{Fig.6}
\end{figure}

In this subsection, we evaluate the performance of the Shapley value-based payoff allocation presented in Proposition~\ref{Proposition1} and the stability condition established in Theorem~\ref{Theorem2}. As illustrated in Fig.~\ref{Fig.6}, we compare the payoffs received by each ET $i\!\in\!\mathcal{N}_e$ and each FPT $i\!\in\!\mathcal{N}_f$ across $14$ scenarios, where the total number of trucks $N$ varies from $2$ to $15$. In each scenario, the number of FPTs varies from $1$ to $(N\!-\!1)$, and truck payoffs are computed with~\eqref{Equ.18}. Default parameter settings are employed, i.e., $\epsilon_f\!=\!0.07$~\texteuro/km and $\epsilon_e\!=\!0.048$~\texteuro/km. Fig.~\ref{Fig.6} shows that the Shapley value-based payoff for each FPT increases with the platoon size, independent of the number of ETs in the platoon. The payoff for each ET increases with the number of FPTs in the formation. This indicates that ETs benefit more from platoons with a higher proportion of FPTs, while FPTs gain more from larger platoons, consistent with our discussions on Proposition~\ref{Proposition1}. Moreover, Fig.~\ref{Fig.6} shows that, for any given $N$, the payoffs for each ET and FPT are identical when $N_e\!=\!1$. As stated in Proposition~\ref{Proposition2}, under this condition, the Shapley value-based allocation coincides with the allocation $x_i$ defined in \eqref{Equ.7}, with $\phi_i\!=\!x_i$ and $\xi\!=\!\frac{1}{N}$.

\begin{figure}[t] 
\centering
\includegraphics[scale=0.44]{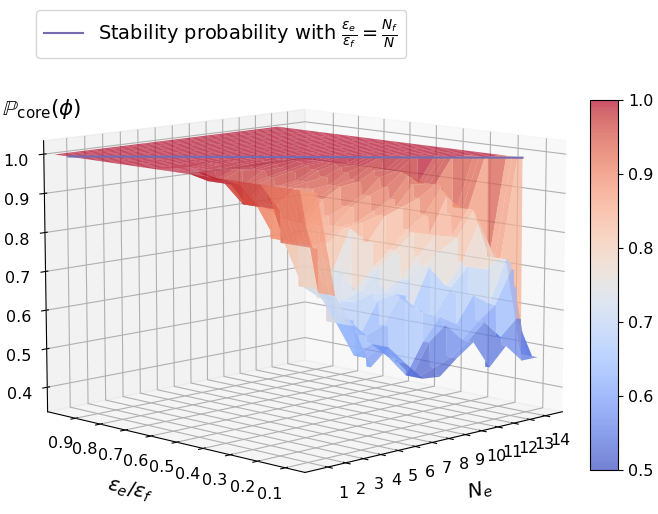}
\vspace{-1pt}
\caption{Stability probability of allocation $\phi$.}
\label{Fig.7}
\end{figure}

\begin{figure}[t] 
\centering
\includegraphics[scale=0.17]{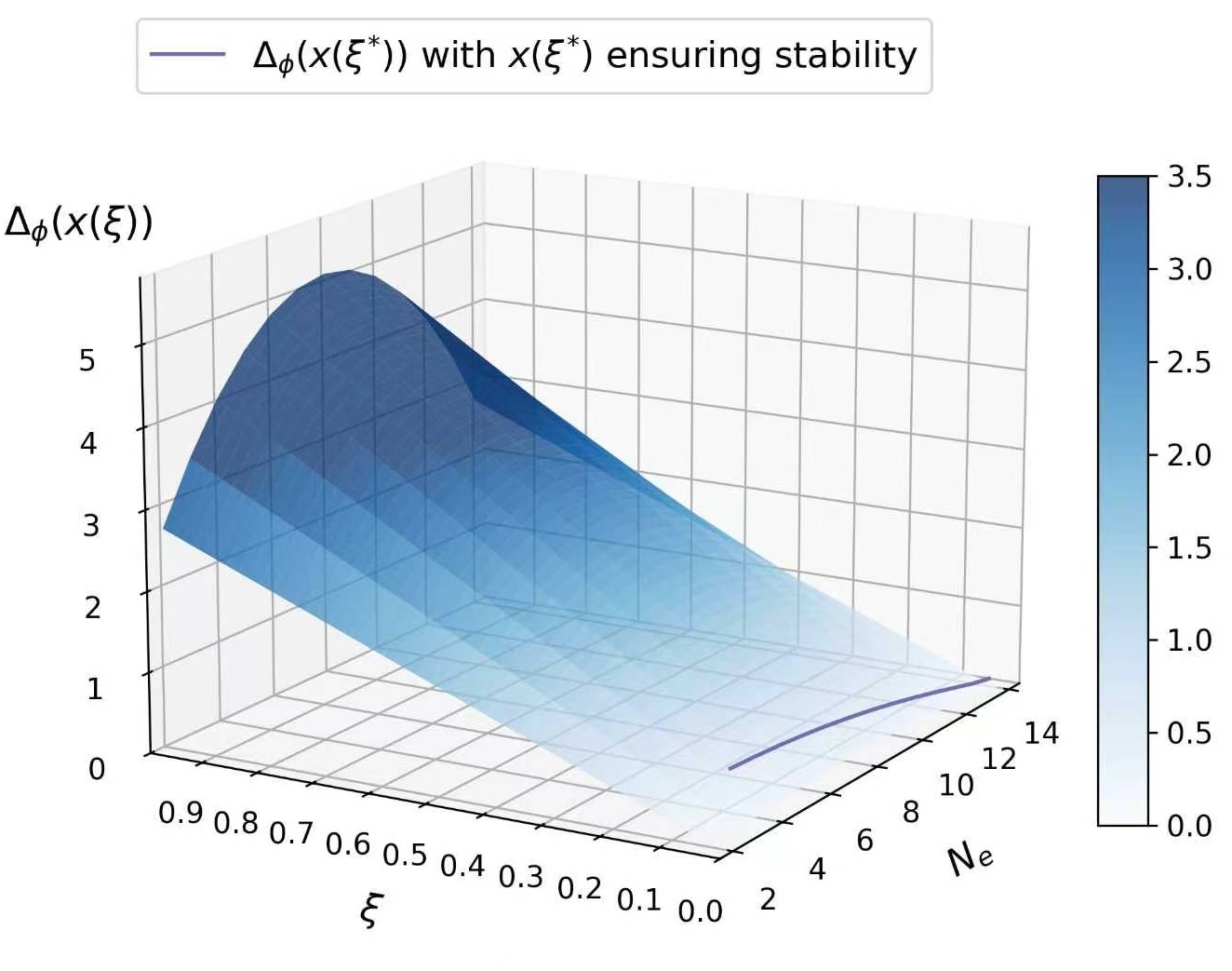}
\vspace{-3pt}
\caption{Mean relative deviation $\Delta_{\phi}(x(\xi))$.}
\label{Fig.8}
\end{figure}

Fig.~\ref{Fig.7} shows the stability probability $\mathbb{P}_{\mathrm{core}}(\phi)$ of the Shapley value-based allocation $\phi$ in different parameter settings, where $\frac{\epsilon_e}{\epsilon_f}\!\in\!(0,1)$ and $N_e$ varies from $1$ to $14$ with $N\!=\!15$. In line with~\eqref{Equ.37}, the stability probability of $\phi$ is denoted as 
\begin{eqnarray}
\mathbb{P}_{\mathrm{core}}(\phi)&&\!=\mathbb{P}\big(\phi\!\in\!\mathrm{Core}(G)\big)\nonumber\\
&&\!=1\!-\!\frac{\big|\{\mathcal{S}\!\subset\!{\mathcal{N}}\!:\!v(\mathcal{S})\!>\!\sum_{i\in{\mathcal{S}}}\!\phi_i,\mathcal{S}\!\neq\!{\emptyset}\}\big|}{2^N\!\!-\!2},\nonumber
\end{eqnarray}
where $\phi_i\!\in\!\phi$ is the Shapley value-based allocation of truck $i\!\in\!\mathcal{N}$ computed by \eqref{Equ.18}. The boundary $\frac{\epsilon_e}{\epsilon_f}=\frac{N_f}{N}$, depicted by the blue line in Fig.~\ref{Fig.7}, represents the threshold beyond which the core stability of $\phi$ is guaranteed. As we can see from the figure, the stability probability $\mathbb{P}_{\mathrm{core}}(\phi)\!=\!1$ when $\frac{\epsilon_e}{\epsilon_f}\!\geq\!{\frac{N_f}{N}}$. This validates the stability condition established in Theorem~\ref{Theorem2}. 

\begin{figure*}[t]
    \centering
    \subfigure[Average stability rate and average $\Delta_{\phi}(x)$.]{
        \includegraphics[width=0.485\linewidth]{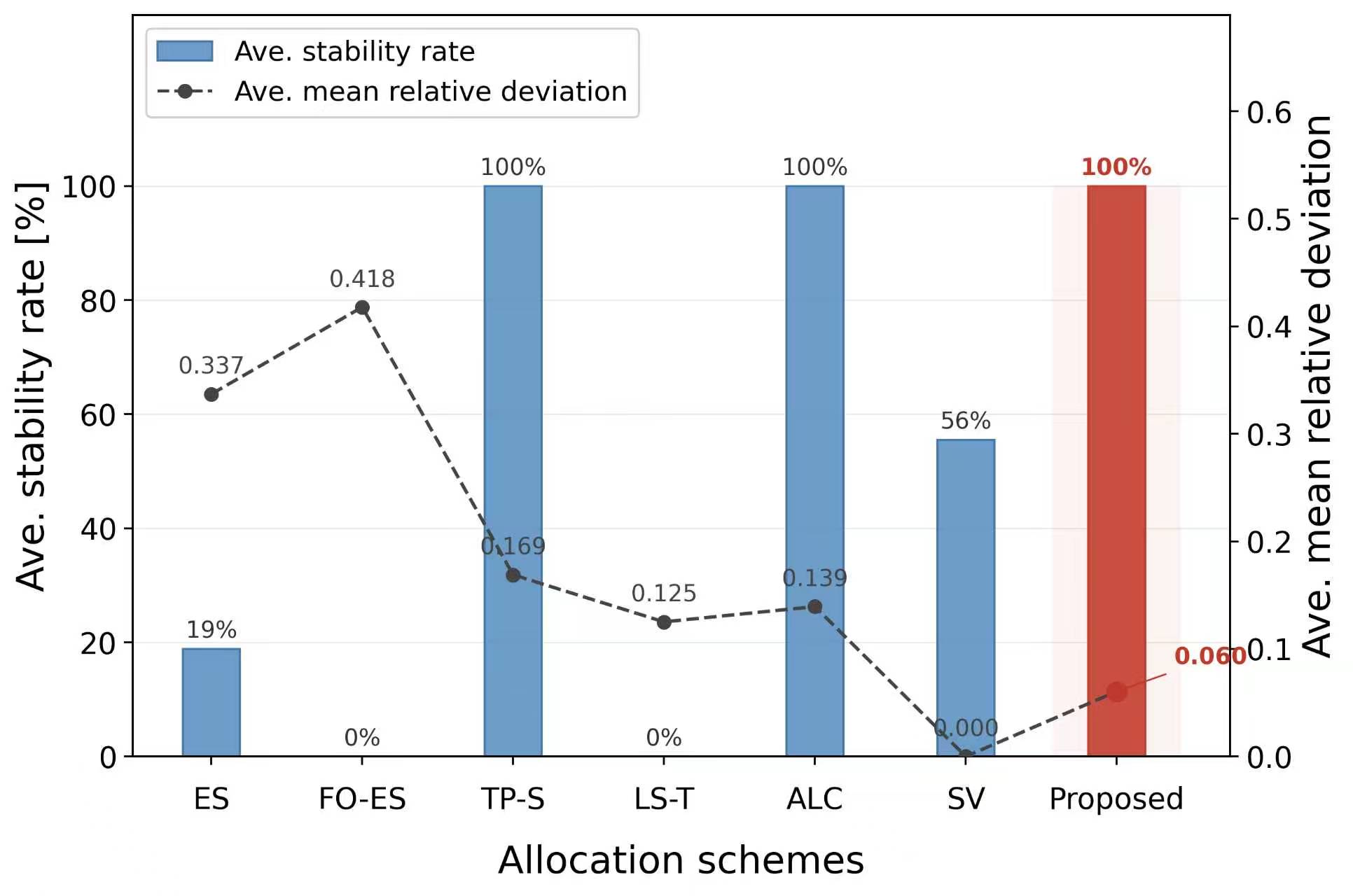}
        \label{Fig.9a}
    }
    \subfigure[Distribution of mean relative deviation.]{
        \includegraphics[width=0.455\linewidth]{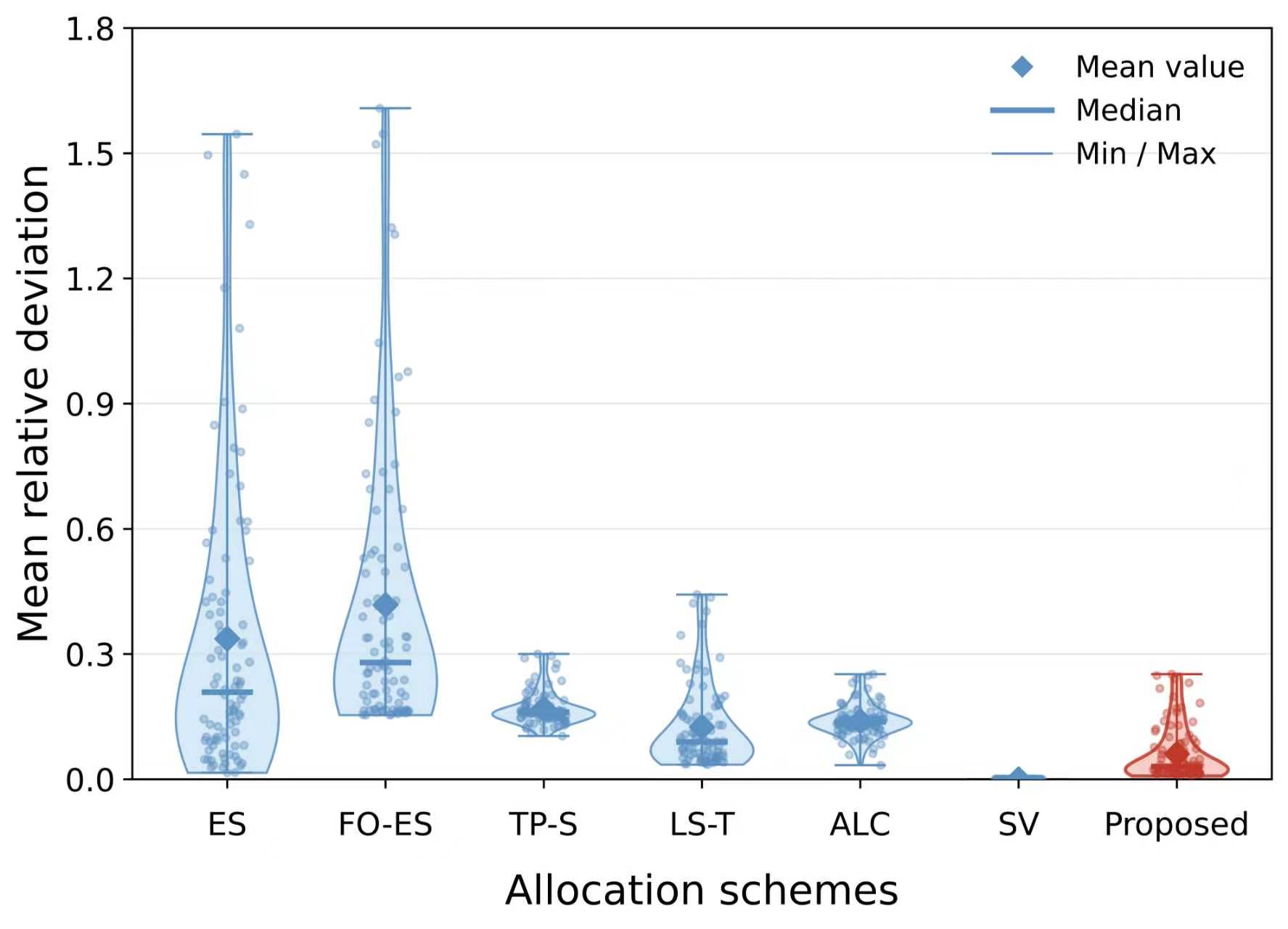}
        \label{Fig.9b}
    }
    \caption{Comparison of core-stability and fairness in different allocation schemes.}
    \label{Fig.9}
\end{figure*}

\subsection{Evaluation of the mean relative deviation}\label{Section6.3}
Next, we evaluate the performance of the payoff allocation scheme proposed in Theorem~\ref{Theorem3}, which is designed for scenarios where the Shapley value-based allocation lies outside the core, i.e., when $\frac{\epsilon_e}{\epsilon_f}\!\geq\!{\frac{N_f}{N}}$ is not satisfied for $N_f\!>\!0$ and $N_e\!>\!{1}$. Fig.~\ref{Fig.8} shows the mean relative deviation $\Delta_{\phi}(x(\xi))$ across different values of $\xi$ over the range $0\!<\!\xi\!<\!{1}$. The number of ETs $N_e$ ranges from $2$ to $14$. The parameters are set to $\epsilon_f\!=\!0.725$~\texteuro/km and $\epsilon_e\!=\!0.048$~\texteuro/km, ensuring that condition \eqref{Equ.28} is violated. As shown by the blue line in Fig.~\ref{Fig.8}, the deviation $\Delta_{\phi}(x(\xi^*))$ remains consistently low across various scenarios and satisfies $\Delta_{\phi}(x(\xi^*))\!<\!1$. This demonstrates the effectiveness of the allocation scheme proposed in Theorem~\ref{Theorem3} in improving fairness while maintaining core stability, even under unfavorable parameter conditions.
\vspace{-5pt}

\subsection{Comparison with baselines}
To further evaluate the practical advantage of the proposed scheme, we compare it with six representative baseline allocation mechanisms, as introduced below.
\vspace{-10pt}
\begin{itemize}
\item [\small{$\bullet$}] \textbf{Equal split (ES)}: The total platooning benefit is shared equally among all trucks, regardless of their role or type.
\vspace{2pt}

\item [\small{$\bullet$}] \textbf{Follower-only equal split (FO-ES)}: The total platooning benefit is equally shared only among the followers, while the leader receives no allocation.
\vspace{2pt}

\item [\small{$\bullet$}] \textbf{Type-proportional sharing (TP-S)}: The platooning benefit is allocated to followers in proportion to their type-dependent saving coefficients $\epsilon_i\!\in\!\{\epsilon_e,\epsilon_f\}$, while the leader receives no allocation. 
\vspace{2pt}

\item [\small{$\bullet$}] \textbf{Leader-subsidy transfer (LS-T)}: The leader receives a fixed fraction $\tau\!\in\!(0,1)$ of the total platooning benefit, and the remaining benefit is distributed to followers according to their saving coefficients. For any truck $i\!\in\!\mathcal{N}$, the allocation is
\vspace{-8pt}
\begin{eqnarray}
x_i\!=\!
\begin{cases}
\!\tau v(\mathcal{N}), & ~\text{if } i\!=\!L,\\
\!\dfrac{\epsilon_i}{\sum_{j\in\mathcal{N}\setminus\{L\}}\!\epsilon_j}(1\!-\!\tau)v(\mathcal{N}), & ~\text{otherwise}.
\end{cases}\nonumber
\end{eqnarray}
\vspace{-8pt}
\item [\small{$\bullet$}] \textbf{Auction-based leader compensation (ALC)}: Each follower $j\!\in\!\mathcal{N}\!\setminus\!\{L\}$ submits a bid $b_j\!=\!\beta\epsilon_j$ with $\beta\!\in\!(0,1)$. The leader is compensated based on the highest bid, and the remaining benefit is allocated to the followers proportionally to their saving coefficients. For any truck $i\!\in\!\mathcal{N}$, the allocation is
\vspace{-8pt}
\begin{eqnarray}
x_i\!=\!
\begin{cases}
\!\beta \max_{j\in\mathcal{N}\setminus\{L\}}\!\epsilon_j, & ~\text{if } i\!=\!L,\\
\!\dfrac{\epsilon_i}{\sum_{j\in\mathcal{N}\setminus\{L\}}\!\epsilon_j}
\big(v(\mathcal N)\!-\!x_L\big), & ~\text{otherwise}.
\end{cases}\nonumber
\end{eqnarray}
\vspace{-8pt}
\item [\small{$\bullet$}] \textbf{Shapley value (SV)}: The benefit is allocated according to each truck’s expected marginal contribution over all possible coalitions, as given in \eqref{Equ.18}. 
\end{itemize}
\vspace{-5pt}

These baselines cover \textit{heuristic allocation rules} (ES, FO-ES, TP-S), \textit{contract-based mechanisms} (LS-T), \textit{auction-based schemes} (ALC), and \textit{marginal contribution-based allocations} (SV). In the simulation, we consider mixed-energy platoons with $N\!=\!12$, $M\!=\!12$, $N_f\!=\!1,\dots,10$, and $\epsilon_e/\epsilon_f\!=\!0.1,\dots,0.9$. For LS-T and ALC, the tuning parameters $\tau$ and $\beta$ are optimized over $\{0.1,\dots,0.9\}$ via a global grid search to improve the baseline performance. The core stability and fairness of each allocation method are evaluated over all tested parameter settings. 
\vspace{-3pt}

Fig.~\ref{Fig.9a} compares the average stability rate and the average mean relative deviation $\Delta_{\phi}(x)$ of different allocation schemes. Here, stability under each parameter setting is defined by a binary indicator which equals $1$ if the allocation lies in the core and $0$ otherwise, and $\Delta_{\phi}(x)$ is defined in \eqref{Equ.29}. The results show that FO-ES and LS-T are unstable, while ES and SV achieve relatively low average stability rates. Although TP-S and ALC achieve $100\%$ core stability, their average mean relative deviations remain larger than that of the proposed method, which is only $0.06$. Fig.~\ref{Fig.9b} further shows the distribution of $\Delta_{\phi}(x)$ over all tested parameter settings. In particular, ES and FO-ES exhibit widely spread deviations, with some values exceeding $1$. By contrast, the proposed scheme achieves the smallest mean relative deviation among all core-stable baselines and shows a more concentrated distribution, indicating more consistent fairness performance. These results together show that the proposed allocation scheme offers the best trade-off between stability and fairness among the baselines.

\section{Conclusion}\label{Section 7}
This paper investigated the benefit allocation problem within a mixed-energy truck platoon, where distinctions arose not only between the platoon leader and followers but also among follower trucks of different types, specifically fuel-powered and electric trucks. We first modeled the interactions among trucks forming platoons as a coalitional game, and proposed a novel benefit allocation scheme with parameter designs that ensure core stability of the payoff distribution. Additionally, we developed a Shapley value–based allocation approach tailored for scenarios where participating trucks prioritize fairness in platoon formation. An efficient closed-form Shapley value was provided, with sufficient conditions guaranteeing both fairness and core stability. For special cases where the Shapley value falls outside the core, we introduced a modified stable payoff allocation that minimizes the mean relative deviation from the Shapley value and proved that the deviation is upper bounded by $1$, achieving minimal unfairness while preserving core stability. Extensive numerical studies verified our theoretical results. In future work, we will incorporate departure time scheduling into the proposed benefit allocation framework to enhance platooning opportunities and maximize the resulting profits.

\bibliographystyle{plainnat}
\bibliography{ting,IDS}          

\end{document}